\documentclass[11pt,dvips,twoside,letterpaper]{article}
\usepackage{graphicx}
\usepackage{times}
\usepackage{fancyhdr}
\usepackage{geometry}
\usepackage{latexsym}
\font\tenfrakturb=eufb10
\font\tenfraktur=eufm10
\font\tenmsbm=msbm10
\font\sevenfrakturb=eufb7
\font\sevenfraktur=eufm7
\font\sevenmsbm=msbm7
\font\fivefrakturb=eufb5
\font\fivefraktur=eufm5
\font\fivemsbm=msbm5
\newfam\bgothicfam
\newfam\gothicfam
\newfam\msbmfam
\textfont\bgothicfam = \tenfrakturb \scriptfont\bgothicfam=\sevenfrakturb
\scriptscriptfont\bgothicfam=\fivefrakturb
\textfont\gothicfam = \tenfraktur \scriptfont\gothicfam=\sevenfraktur
\scriptscriptfont\gothicfam=\fivefraktur
\textfont\msbmfam = \tenmsbm \scriptfont\msbmfam=\sevenmsbm
\scriptscriptfont\msbmfam=\fivemsbm

\def\Bbb{\tenmsbm\fam\msbmfam}

\catcode`@=11
\def\renewcounter#1{\@definecounter{#1}\@ifnextchar[{\@newctr{#1}}{}}
\begin{document}
\title{
~\\[-1.2in]
{\normalsize\noindent
\begin{picture}(0,0)(208,-3.15)
%\begin{tabular}{l p{0.65in} r}
\begin{tabular}{l p{0.7in} r}
In: New Developments in Black Hole Research & & ISBN: 1-59454-641-X \\
Editor: Paul V. Kreitler  & & \copyright 2005 Nova Science Publishers, Inc.\\
%Journal  & & ISSN xxxxxxxxxx \\
%Volume 11, Number 4, pp. 00-00  & & \copyright~2005 Nova Science Publishers, Inc.\\
\end{tabular}
\end{picture}}
{\begin{flushleft} ~\\[0.63in]{\normalsize\bfseries\textit{Chapter~3}} \end{flushleft} ~\\[0.13in] 
\bfseries\scshape Intersection of Black Hole Theory and Quantum 
Chromodynamics: The Gluon Propagator Corresponding to Linear Confinement at Large Distances and Relativistic Bound States in the Confining
SU($N$)-Yang-Mills Fields}}
\author{
\bfseries\itshape Yu. P. Goncharov\\
%\thanks{E-mail address: xxxx}\\
Theoretical Group, Experimental Physics Department, State\\
Polytechnical University, Sankt-Petersburg 195251, 
Russia}
\date{}
\maketitle
\thispagestyle{empty}
\setcounter{page}{1}

% ------------ Running Heads -------------------------------------------------
\pagestyle{fancy}
%\fancyhf{}
\fancyhead{}
\fancyhead[EC]{Yu. P. Goncharov}
\fancyhead[EL,OR]{\thepage}
\fancyhead[OC]{Intersection of Black Hole Theory and Quantum 
Chromodynamics...}
\fancyfoot{}
\renewcommand\headrulewidth{0.5pt}
\addtolength{\headheight}{2pt} % make space for the rule
\headsep=9pt
%-------------------------------------------------------------------------------
\begin{abstract}
The exact nonperturbative confining solutions of the SU(3)-Yang-Mills equations 
recently obtained by author in Minkowski spacetime with 
the help of the black hole theory techniques are analysed and on the basis
of them the gluon propagator corresponding to linear confinement at large distances 
(small momenta) is constructed in a nonperturbative way. At small distances 
(large momenta) the resulting 
propagator passes on to the standard (nonperturbative) gluon propagator used 
in the perturbative quantum chromodynamics (QCD). The results suggest some 
scenario of linear 
confinement for mesons and quarkonia which is also outlined. As a consequence 
there arises a motivation for studying the relativistic bound states in the 
above confining SU($N$)-Yang-Mills fields. This possiblity is realized for 
$N=2,3,4$ with the aid of the black hole theory results about spinor fields on 
black holes with a subsequent application to the charmonium spectrum in the 
most important physical case $N=3$. Incidentally uniqueness of the confining 
solutions is discussed and a comparison with the nonrelativistic potential 
approach is given. 
\end{abstract}
%\newpage
\tableofcontents
\section{Introduction and Preliminary Remarks}
As soon as quantum chromodynamics (QCD) was proposed as the main candidate
for the theory of strong interactions \cite{Q} at once there arised the
question about the confinement of quarks within the framework of QCD. In the 
late seventies of XX century the main approaches to solve the problem were
formed (see, e.g., review of Ref. \cite{Ban81}) and they actually remain the 
same ones up to now. For this purpose miscellaneous
techniques were elaborated, for example, strong coupling expansions, lattice
approach, instanton improvement of perturbation theory, nonrelativistic 
potential approach and so on. It should
be noted, however, none of the mentioned various directions has so far led to 
a generally accepted theory of quark confinement.

  In this paper we would like from another side to have analysed one of the
possible approaches. The question is about a nonperturbative modification of 
gluon propagator which might correspond to linear confinement between quarks
at large distances. The very simple idea of modifying the mentioned propagator
arises when considering the naive Fourier transform for the power potentials of 
form $r^\lambda$ (for more details, see e.g. Ref. \cite{Ban81}). Then at
$\lambda=1$ (linear confinement) the conforming Fourier transform (propagator) 
is of order $|{\bf k}|^{-4}$ in momentum space, while the case 
$\lambda=-1$ (Coulomb potential) gives the standard gluon propagator 
$\sim |{\bf k}|^{-2}$. All the attempts to obtain the necessary behaviour, 
however, for example, by summing a selective infinite set of perturbation 
diagrams with using the Dyson--Schwinger equations for the propagator 
failed \cite{Ban81}. It is clear why: it is impossible to get anything 
nonperturbative like confinement by perturbative techniques. Some new 
possibilities in this direction were connected 
with lattice theories (for more details, see, e.g., Ref. \cite{Lat02} and 
references therein) but the results here are mainly of qualitative character.

 To our mind, from the very outset the problem should be considered
on the basis of the exact nonperturbative solutions of the SU(3)-Yang-Mills 
equations modelling quark confinement which, in what follows, we shall call 
{\em the confining solutions}. Such solutions will be 
supposed to contain only the components of the SU(3)-field which are 
Coulomb-like or linear in $r$, the distance between quarks. In 
Ref. \cite{Gon01} a number of such solutions have been obtained and the 
corresponding spectrum of Dirac equation describing the relativistic bound
states in those confining SU(3)-Yang-Mills fields has been analysed. Further
in Refs. \cite{{Gon03},{GC03},{Gon04}} the results obtained were successfully applied 
to the description of the quarkonia spectra (charmonium and bottomonium). In 
its turn, the mentioned description suggests that linear confinement is 
(classically) governed by the magnetic colour field linear in $r$ and, 
as was mentioned in Refs. \cite{{Gon03},{Gon05}},  
one can try to modify the gluon propagator nonperturbatively at quantum level 
for to generate the mentioned magnetic colour field at classical level. One part of 
the present paper is just devoted to it. It should be noted, however, that all the main 
features of such a modification may occur already within quantum 
electrodynamics (QED) that should be not surprising because, as is historically 
known (see, e.g., Ref. \cite{Ryd85}), the standard gluon propagator is in fact 
the slightly modified photon one of QED. Under the circumstances we shall 
conduct our considerations mainly within QED but incidentally making remarks 
to generalize the results obtained to the QCD case. Considerations of the first 
part of paper (Sections 1--4) inevitably lead to the task of a more thorough 
analysis of the Yang--Mills and Dirac equations derived from QCD lagrangian 
which is realized in the rest of paper.

     Let us introduce some notations. Further we shall deal with the metric of
the flat Minkowski spacetime $M$ that
we write down [using the ordinary set of local rectangular (Cartesian) 
($x,y,z$) or spherical ($r,\vartheta,\varphi$) coordinates
for spatial part] in the forms
$$ds^2=g_{\mu\nu}dx^\mu\otimes dx^\nu\equiv
dt^2-dx^2-dy^2-dz^2\equiv
dt^2-dr^2-r^2(d\vartheta^2+\sin^2\vartheta d\varphi^2)\>, \eqno(1.1)$$
so the components $g_{\mu\nu}$ take different values depending on the
choice of coordinates.
Besides we have $\delta=|\det(g_{\mu\nu})|=(r^2\sin\vartheta)^2$
in spherical coordinates and the exterior differential $d=\partial_t dt+
\partial_xdx+\partial_ydy+\partial_zdz$ or $d=\partial_t dt+\partial_r dr+
\partial_\vartheta d\vartheta+\partial_\varphi d\varphi$ in the corresponding
coordinates. We denote 3-dimensional vectors by bold font
so $X=(t,x,y,z)\equiv(t,{\bf r}), k=(k_0,k_1,k_2,k_3)\equiv(k_0,{\bf k})$ 
with $r^2=x^2+y^2+z^2,|{\bf k}|=\sqrt{k_1^2+k_2^2+k_3^2}$. The Fourier
transform $\tilde{\Phi}(k)$ of some function $\Phi(X)$ is formally defined 
by the relations ($kX=k_0t-k_1x-k_2y-k_3z$)
$$\tilde{\Phi}(k)=\int_M\exp(ikX)\Phi(X)d^4X=F[\Phi]\>,$$
$$\Phi(X)=\frac{1}{(2\pi)^4}\int_M\exp(-ikX)\tilde{\Phi}(k)d^4k\>\eqno(1.2)$$
but it is treated in the sense of the theory of generalized functions 
(distributions) (see, e.g., Refs. \cite{GS}) and we denote 
$d^4k=dk_0dk_1dk_2dk_3$,
$d^4X=dtdV$, $dV=dxdydz$ or $dV=\sqrt{\delta}drd\vartheta d\varphi$ while 
for the generalized $\delta$-functions we use the notations 
$\delta(X)\equiv\delta(t)\delta(x)\delta(y)\delta(z)$,
$\delta({\bf r})\equiv\delta(x)\delta(y)\delta(z)$. Other mathematical results 
necessary for our considerations are gathered in Appendices $A,B,C,D,E$. 

  Throughout the paper we employ the system of units with $\hbar=c=1$,
unless explicitly stated otherwise. Finally, we shall denote $L_2(F)$ the 
set of the modulo square integrable complex functions on any manifold $F$ 
furnished with an integration measure while $L^n_2(F)$ will be the $n$-fold 
direct product of $L_2(F)$ endowed with the obvious scalar product.

\section{Confining Solutions of the Maxwell and SU(3)-Yang-Mills\\
Equations}
\subsection{Black Hole Theory Techniques}
For obtaining a set of the confining solutions within the given paper we shall 
employ the techniques used in Refs. \cite{Gon678} for finding the 
U($N$)-monopole solutions in black hole physics and the essence of those 
techniques consists 
in systematic usage of the Hodge star operator (see Appendix $A$) conforming to 
metric (1.1). As is known, such a metric can be obtained from the Schwarzschild 
black hole metric when the black hole mass is equal to 0.

Really, if writing down the Yang-Mills equations ($B.3$) in components then we 
shall be drowned in a sea of indices which will strongly hamper searching for one 
or another ansatz and make it practically immense. Using the Hodge star 
operator as well as the rules of external calculus makes the problem  
quite foreseeable and quickly leads to the aim. 

As was remarked in Appendix $B$, the sought solutions are usually 
believed to obey an additional condition and as the latter one 
in the present paper we take the Lorentz condition that can be written
in the form 
$${\rm div}(A)=0\>, \eqno(2.1)$$ 
where the divergence of the Lie algebra valued
1-form $A=A^a_\mu T_adx^\mu$ is defined by the relation 
$${\rm div}(A)=\frac{1}{\sqrt{\delta}}\partial_\mu(\sqrt{\delta}g^{\mu\nu}
A_\nu)\>.\eqno(2.2)$$

\subsection{Electrodynamics}
 We proceed from the second pair of Maxwell equations ($B.5$) 
$$d\ast F= J\>\eqno(2.3)$$
with $F=dA$, $A=A_\mu dx^\mu$ and the Hodge star operator $*$ is defined,
for example, on 2-forms $F=F_{\mu\nu}dx^\mu\wedge dx^{\nu}$ in Minkowski 
spacetime $M$ provided with a pseudoriemannian metric $g_{\mu\nu}$ (1.1) by 
the relation (see Appendix $A$)
$$
F\wedge\ast F=(g^{\mu\alpha}g^{\nu\beta}-g^{\mu\beta}g^{\nu\alpha})
F^a_{\mu\nu}F^a_{\alpha\beta}
\sqrt{\delta}\,dx^1\wedge dx^2\cdots\wedge dx^4 
\eqno(2.4)
$$
in local coordinates $x=(x^\mu)$ while $J=j_\mu*(dx^\mu)$ with a 4-dimensional
electromagnetic density current $j=j_\mu dx^\mu$. Let $J=0$ and we shall search for 
the solution of (2.3) in the form $A=A_t(r)dt+A_\varphi(r)d\varphi$.
It is then easy to check that $F=dA=
-\partial_rA_tdt\wedge dr+\partial_rA_\varphi dr\wedge d\varphi$ and
since $\ast(dt\wedge dr)=-r^2\sin\vartheta d\vartheta\wedge d\varphi$,
$\ast(dr\wedge d\varphi)=-\frac{1}{\sin\vartheta}dt\wedge d\vartheta$ we get
$\ast F=r^2\sin\vartheta\partial_rA_td\vartheta\wedge d\varphi-
\frac{1}{\sin\vartheta}\partial_rA_\varphi dt\wedge d\vartheta$. From here
it follows that Eq. (2.3) yields
$$\partial_r(r^2\partial_rA_t)=0,\>\partial^2_rA_\varphi=0\>,\eqno(2.5)$$
and we write down the solutions of (2.5) as 
$$ A_t =\frac{a}{r}+A \>, A_\varphi=br+B \>\eqno(2.6)$$
with some constants $a, b, A, B$ parametrizing solutions (further for the 
sake of simplicity let us put $a=1, b=1\ {\rm GeV}, A=B=0$).
 
To interpret solutions (2.6) in the more habitual physical terms let us pass on
to Cartesian coordinates employing the relations 
$$\varphi=\arctan(y/x),\>
d\varphi=\frac{\partial\varphi}{\partial x}dx+
\frac{\partial\varphi}{\partial y}dy\> \eqno(2.7)$$
which entails
$$A_\varphi d\varphi=-\frac{ry}{x^2+y^2}dx+\frac{rx}{x^2+y^2}dy\>\eqno(2.8)$$
and we conclude that the solutions of (2.6) describe the combination
of the electric Coulomb field with potential $\Phi=1/r$ and the constant 
magnetic field with vector-potential
$${\bf A}=(A_x,A_y,A_z)=\left(-\frac{ry}{x^2+y^2},\frac{rx}{x^2+y^2},0\right)=
\left(-\frac{\sin\varphi}{\sin\vartheta},\frac{\cos\varphi}{\sin\vartheta},
0\right)\>,
\eqno(2.9)$$
which is {\it linear} in $r$ in spherical coordinates and the 3-dimensional 
divergence ${\rm div}{\bf A}=0$, as can be checked directly. Then Eq. (2.3) in 
Cartesian coordinates takes the form
$$\Delta\Phi=0,\> {\rm rotrot}{\bf A}= \Delta{\bf A}=0\> \eqno(2.10)$$
with the Laplace operator $\Delta=\partial_x^2+\partial_y^2+\partial_z^2$.
At last it is easy to check that the solution under consideration satisfies the 
Lorentz condition (2.2)
$${\rm div}(A)=\frac{1}{\sqrt{\delta}}\partial_\mu(\sqrt{\delta}g^{\mu\nu}
A_\nu)=0\>.\eqno(2.11)$$

\subsection{SU(3)-Yang-Mills Theory}
 Now Eq. (2.3) should be replaced by the Yang-Mills equations ($B.3$)
$$d\ast F= g(\ast F\wedge A - A\wedge\ast F) +J\>\eqno(2.12)$$
for SU(3)-field $A=A_\mu dx^\mu$, $A_\mu=A^a_\mu T_a$ where the matrices 
$T_a$ form a basis of the Lie algebra of SU(3) in 3-dimensional space, 
$a=1,...,8$ and 
further let us put $T_a=\lambda_a$, where $\lambda_a$ are the Gell-Mann 
matrices (see Appendix $B$).
After this we search (at $J=0$) for the solution of (2.12) in the form 
$A=A_t(r)dt+A_\varphi(r)d\varphi$
with $A_{t,\varphi}=A^3_{t,\varphi}\lambda_3+A^8_{t,\varphi}\lambda_8$.
Evaluating $dF=dA+gA\wedge A$ it is easy to check that the right-hand side of
(2.12) is equal to zero and to gain the
sought solution in the form (which reflects the fact that for any matrix 
${\cal T}$ from SU(3)-Lie algebra we have ${\rm Tr}\,{\cal T}=0$) 
 $$ A^3_t+\frac{1}{\sqrt{3}}A^8_t =-\frac{a_1}{r}+A_1 \>,
 -A^3_t+\frac{1}{\sqrt{3}}A^8_t=-\frac{a_2}{r}+A_2\>,
-\frac{2}{\sqrt{3}}A^8_t=\frac{a_1+a_2}{r}-(A_1+A_2)\>, $$
$$ A^3_\varphi+\frac{1}{\sqrt{3}}A^8_\varphi =b_1r+B_1 \>,
 -A^3_\varphi+\frac{1}{\sqrt{3}}A^8_\varphi=b_2r+B_2\>,
-\frac{2}{\sqrt{3}}A^8_\varphi=-(b_1+b_2)r-(B_1+B_2)\>, \eqno(2.13)$$
where real constants $a_j, A_j, b_j, B_j$
parametrize the solution, and we wrote down
the solution in the combinations that are just
needed to insert into the corresponding Dirac 
equation (see Section 5). From here it follows
$$A^3_t = [(a_2-a_1)/r+A_1-A_2]/2,\>
A^8_t =[A_1+A_2-(a_1+a_2)/r]\sqrt{3}/2\>,$$
$$ A^3_\varphi = [(b_1-b_2)r+B_1-B_2]/2,
   A^8_\varphi= [(b_1+b_2)r+B_1+B_2]\sqrt{3}/2\>\eqno(2.14)$$
and practically the same considerations as the above ones in electrodynamics
show that the given solution describes the configuration of the electric 
Coulomb-like colour field (components $A_t$) with potentials $\Phi^3,\Phi^8$ 
and the constant magnetic colour 
field (components $A_\varphi$) with vector-potentials ${\bf A}^3,{\bf A}^8$
which are {\it linear} in $r$ in spherical coordinates with 3-dimensional 
divergences ${\rm div}{\bf A^3}$=${\rm div}{\bf A^8}=0$ while the Eq. (2.12)
is easily transformed into the Eqs. (2.10) with an obvious modification.
It is also simple to check that the solution under consideration satisfies the 
Lorentz condition (2.2).

\section{Linear Confinement in QED}
Now we can investigate how the photon propagator should be 
modified if in the real world the interaction between two charged particles
would not be classically described only by the Coulomb law but it would also 
include a constant magnetic field linear in $r$, the distance between particles,
so the given field would obey the Maxwell equations.
Let us briefly recall the scheme in accordance with that the standard photon 
propagator is obtained (see, e.g., Ref. \cite{Ryd85}).
\subsection{Standard Photon Propagator}
 In Cartesian coordinates the Eq. (2.3) for $A=A_\mu dx^\mu$ at $J=0$ with 
Lorentz condition ${\rm div}(A)=0$ reduces to the system
$$\Box A_\mu =0 \> \eqno(3.1)$$
with the d'Alembert operator $\Box=\partial_t^2-\Delta$. 
Then one constructs the fundamental solution (a Green function) of the 
system (3.1) (the photon propagator at classical level) as the matrix
with elements $D_{\mu\nu}(X)$ obeying the system
$$g_{\sigma\mu}\Box D_{\mu\nu}(X)=-g_{\sigma\nu}\delta(X)\>.\eqno(3.2)$$
Further one uses the
fundamental solution of the d'Alembert operator [i.e., the solution of
the equation $\Box K=\delta(X)$] in the form
$$K(X)=\frac{1}{4\pi^2i(X^2-i0)}\> \eqno(3.3)$$
with quadratic form $X^2=t^2-r^2$ (the exact definitions concerning the 
generalized functions connected with quadratic forms can be found in 
Refs. \cite{GS}), so that
$$D_{\mu\nu}(X)=- g_{\mu\nu}K(X)=- \frac{g_{\mu\nu}}{4\pi^2i(X^2-i0)}\>.
\eqno(3.4)$$
 After this the photon propagator at quantum level is obtained as the Fourier
transform for $D_{\mu\nu}(X)$, namely
$$\tilde{D}_{\mu\nu}(k)=F[D_{\mu\nu}(X)]=- \frac{g_{\mu\nu}}{k^2+i0}\>,
\eqno(3.5)$$
since $\tilde{K}=F[K]=1/(k^2+i0)$ with quadratic form 
$k^2=k_0^2-k_1^2-k_2^2-k_3^2$ (for more details see Refs. \cite{GS}).
There arises the question why among a large set of the mathematically possible 
fundamental solutions (Green functions) for Eq. (3.2) one chooses just the 
one of (3.4). The 
answer can be based only on physical considerations. 

Indeed, let us take a point particle 
with a charge $e$ moving with a velocity ${\bf v}={\bf v}(t)$. Then, as is 
known (see, e.g. Ref. \cite{LL}), the 4-dimensional density current of such
an object is
$j=j_\mu dx^\mu=e\delta({\bf r})(dt+{\bf v}d{\bf r})$,
$d{\bf r}=(dx,dy,dz)$. Under the circumstances the particle will generate
the electromagnetic field with potential $A=A_\mu dx^\mu$ which 
should be obtained by the contraction of $j$ with a fundamental solution 
$D_{\mu\nu}(X)$ of Eq. (2.3) or, that is the same, of Eq. (3.1), namely 
$$A_\mu(X)=4\pi\int D_{\mu\nu}(X-X')j_\nu(X')d^4X' \>.\eqno(3.6)$$
When choosing $D_{\mu\nu}(X)$ equal to that of (3.4) we obtain, for example,
for electric potential of the field generated [with replacing $t\to it$ in the 
integral over $t$ in (3.6)]
$$A_t(X)=\Phi(X)= \frac{e}{\pi}\int_{-\infty}^\infty\frac{dt}{t^2+r^2}=
\frac{e}{r}\>, \eqno(3.7)$$
that is, the Coulomb law. Analogously, other diagonal components of 
$D_{\mu\nu}(X)$ of (3.4) give, for example, at ${\bf v}=const$ the 
vector-potential ${\bf A}$ for a Coulomb-like magnetic field generated by the 
point charged particle when its moving (for more details see Ref. \cite{LL}).
All of that corresponds to experimental data and, as a result, the choice of
$D_{\mu\nu}(X)$ in the form (3.4) reflects the real situation in our world.

\subsection{Electrodynamics with Linear Confinement}
Let us now explore how the photon propagator should be modified if 
experiment would say to us that when its moving a point charged particle 
generates a constant magnetic field linear in $r$, the distance from the 
particle, additionally to the mentioned Coulomb-like fields.

  Under the circumstances we should use the property of any fundamental solution
that the latter is determined only to within adding any solution of the
conforming homogeneous equation. As we have seen above, components of ${\bf A}$
from (2.9) are the solutions of the Laplace equation. On the other hand, as is 
known, the Coulomb potential is also the solution of the Laplace equation at 
$r\ne0$ and besides
it is a fundamental solution of the Laplace operator \cite{GS}. Consequently,
for to obtain the necessary modification of 3-dimensional photon propagator 
we should add components $A_x$ or $A_y$ of (2.9) to the Coulomb fundamental 
solution for the conforming components of propagator. To pass on
to a 4-dimensional propagator let us recall that the Coulomb fundamental 
solution of the Laplace operator and the fundamental solution (3.3) of the 
d'Alembert operator are connected by the so-called method of descent 
(see, e.g., Ref. \cite{V76}) which is in essence expressed by the relation
$$4\pi\int_{-\infty}^\infty K(X)dt=-\frac{1}{r}\>,\eqno(3.8)$$
so that to obtain the necessary modification of 4-dimensional propagator we 
should add some suitable solutions of the d'Alembert (wave) equation 
$\Box f(X)=0$ to $K(X)$ for the corresponding components of propagator (3.4).
As the latter ones we should take functions $(tA_x)/(8\pi^2)$ or 
$(tA_y)/(8\pi^2)$ [where the factor $1/(8\pi^2)$ is introduced for the sake 
of further convenience] with $A_x$ or $A_y$ of (2.9). The given functions
obviously satisfy the d'Alembert equation. Then the sought photon propagator
will have the same components as in (3.4) except for the cases $\mu=\nu= x$ 
or $y$ where the components will be
$$ D_{xx}(X)=\frac{1}{4\pi^2i(X^2-i0)}+\frac{tA_x}{8\pi^2}=\frac{1}
{4\pi^2i(X^2-i0)}-\frac{try}{8\pi^2(x^2+y^2)} \>,$$
$$ D_{yy}(X)=\frac{1}{4\pi^2i(X^2-i0)}+\frac{tA_y}{8\pi^2}=\frac{1}
{4\pi^2i(X^2-i0)}+\frac{trx}{8\pi^2(x^2+y^2)}\>.\eqno(3.9)$$
Under this situation when its moving a charged particle might generate
an additional constant magnetic field according to the relation (3.6). To 
specify it let us recall that in electrodynamics \cite{LL} any constant 
magnetic field
is connected with a finite motion of charged particles. Let us suppose, 
for instance, that the particle accomplishes a finite motion within a finite
(though perhaps large enough) time in such a way that the velocity 
projections $v_{x,y}(t)$ are some odd functions of time. Then one may consider 
that
$$\int (t-t')v_{x,y}(t')dt'\sim\int t'v_{x,y}(t')dt'\sim C=const\>,
\eqno(3.10)$$
and according to (3.6) there appears some constant magnetic field 
$\sim {\bf A}$ of (2.9) linear in $r$, the distance from particle. So indeed
under the certain conditions we could observe the mentioned magnetic field
corresponding to the propagator described above.
\subsection{Momentum Representation}
To get the necessary propagator at quantum level we should carry out the 
Fourier transform of the just found propagator. So long as for any natural $m$
we have (see Refs. \cite{GS})
$$F[t^m]=2\pi(-i)^m\delta^{(m)}(k_0)\>\eqno(3.11)$$
with $m$-th derivarive of $\delta$-function, then really everything reduces to 
the Fourier transforms for functions $A_x,A_y$ of (2.9). Since the latter ones 
are only locally integrable the Fourier transforms should be understood in the 
sense of the theory of generalized functions \cite{GS}, namely, through 
analytical continuation of suitable integrals. Let us find, e.g., $F[A_x]$.
In accordance with (1.2) we shall have
$$F[A_x]=\int A_x\exp(-i{\bf k}{\bf r})dV=$$
$$-\int_0^\infty r^2dr\int_0^\pi
\exp(-ik_3r\cos\vartheta)d\vartheta\int_0^{2\pi}
\exp[-ir\sin\vartheta(k_1\cos\varphi+k_2\sin\varphi)]\sin\varphi d\varphi
\>.\eqno(3.12)$$
Using the relation of Ref. \cite{PBM1} 
$$\int_0^{2\pi}\exp(a\cos x+b\sin x)\pmatrix{\sin x\cr\cos x\cr}dx=
\frac{\pi}{\sqrt{a^2+b^2}}I_1(\sqrt{a^2+b^2})\pmatrix{2b\cr2a\cr}\>
\eqno(3.13)$$
with the modified Bessel function $I_1(z)=-iJ_1(iz)$,where $J_1(z)=-J_1(-z)$ 
is the standard Bessel function, we can rewrite (3.12) as
$$F[A_x]=\frac{2\pi ik_2}{\sqrt{k_1^2+k_2^2}}\int_0^\infty r^2dr
\int_0^\pi\exp(-ik_3r\cos\vartheta)
J_1(r\sin\vartheta\sqrt{k_1^2+k_2^2})d\vartheta\>.\eqno(3.14)$$
Further replacing $\cos\vartheta=x$ and employing the formula of 
Ref. \cite{PBM2}
$$\int_0^a\frac{\cos(b\sqrt{a^2-x^2})}{\sqrt{a^2-x^2}}J_1(cx)dx=
\frac{1}{ac}[\cos(ab)-\cos(a\sqrt{b^2+c^2})]\>,\eqno(3.15)$$
we get
$$F[A_x]=\frac{4\pi ik_2}{k_1^2+k_2^2}\int_0^\infty r[\cos(k_3r)-
\cos(r|{\bf k}|)]dr\>.\eqno(3.16)$$
At last, using the relation with the Euler $\Gamma$-function of 
Ref. \cite{PBM1} 
$$\int_0^\infty x^{q-1}\cos(mx)dx=
\frac{\Gamma(q)}{m^q}\cos\frac{\pi q}{2}\>,\eqno(3.17)$$
holding true at $0<q<1, m>0 $, we analytically continue the right-hand side
of (3.17) over all admissible values $q,m$ which permits 
[at $q=2$ with $\Gamma(2)=1$] to write down
$$F[A_x]=-\frac{4\pi ik_2}{k_1^2+k_2^2}
\left[\frac{1}{k_3^2}-\frac{1}{|{\bf k}|^2}\right]=
-\frac{4\pi ik_2}{|{\bf k}|^2k_3^2}\>.\eqno(3.18)$$
Analogous consideration yields
$$F[A_y]=\frac{4\pi ik_1}{|{\bf k}|^2k_3^2}\>.\eqno(3.19)$$
After this, employing the relations (3.9) and (3.11) we finally come to the 
conclusion that the sought modification of photon propagator in momentum space 
will have the same components as in (3.5) except for the cases $\mu=\nu=1$ 
or $2$ where the components will be 
$$\tilde{D}_{xx}(k)=\frac{1}{k^2+i0}-
\frac{k_2\delta^{'}(k_0)}{|{\bf k}|^2k_3^2}\>,$$
$$\tilde{D}_{yy}(k)=\frac{1}{k^2+i0}+
\frac{k_1\delta^{'}(k_0)}{|{\bf k}|^2k_3^2}\>,
\eqno(3.20)$$
and the generalized function $\delta^{'}(k_0)$ acts according to the rule
$$\int \Phi(k_0)\delta^{'}(k_0)dk_0 =-\Phi'(0)\>.$$

\section{Linear Confinement in QCD}
\subsection{Motivation}
  It is clear that considerations of the previous section are not confirmed
experimentally in electrodynamics -- there exist no elementary charged 
particles generating a constant magnetic field linear in $r$, the distance 
from particle, the given field obeying the Maxwell equations.

Another matter is quantum chromodynamics. The analogue of charge here is 
colour. The group U(1) and the Maxwell equations are replaced by SU(3) and the 
Yang-Mills equations but, as we have seen in Section 2, both the Maxwell 
equations and the Yang-Mills ones possess the confining solutions. Though
quarks can unlikely be considered classical particles, after all,
they accomplish a finite motion within 
a region with character size of order 1 ${\rm fm}=10^{-13}$ cm and, 
as the explicit form of modulo square integrable solutions of the Dirac equation 
(5.4) in the confining SU(3)-field (2.13)--(2.14) 
(relativistic bound states) shows 
(see Refs. \cite{Gon01,{Gon03},{GC03},{Gon04}} and Section 7), 
the $j$-th colour component for the system of two quarks (e.g., for 
quarkonia) $\Psi_j\sim r^{\alpha_j}e^{-\beta_j r}$ with $\alpha_j>0$, 
$\beta_j=\sqrt{\mu_0^2-\omega_j^2+g^2b_j^2}>0$, where $\omega=\sum\omega_j$ is 
an energetic level of system, $b_{1,2}$ are the parameters of linear interaction 
from the solutions (2.13)--(2.14), $b_3=-(b_1+b_2)$, $r$ is a 
distance between quarks and $\Psi_j$ is proved to be markedly different from 
zero only at $r\sim1/\beta_j\sim$ 0.04 fm (see 
Refs. \cite{{Gon03},{GC03},{Gon04}} and Section 9), 
i.e., we deal with linear confinement of colour. Just the magnetic colour  
field defines the latter through the coefficients $\beta_j$. 
As a result, there are certain 
grounds to consider the qualitative physical picture from the previous section 
to occur just
within QCD and the gluon propagator should be modified. The necessary 
modification can be realized in the same way as is done when deriving
the standard gluon propagator (see, e.g., Ref. \cite{Ryd85}), i.e., through 
multiplying the propagator (3.5) [where the modification (3.20) is implied]
by the factor $\delta^{ab}$ with $a,b=1,...,8$
$$\tilde{D}^{ab}_{\mu\nu}(k)=\delta^{ab}\tilde{D}_{\mu\nu}(k)\>.
\eqno(4.1)$$

  Under the cicumstances the gluon propagator obtained will be able to lead to 
linear confinement at large distances 
(small momenta) while at small distances (large momenta) we can omit the 
additional addenda of order $|{\bf k}|^3$ in (3.20) and the resulting 
propagator will pass on to the standard gluon one used in the 
perturbative QCD.

 It should be noted that during all the considerations we in fact dealt with
the so-called Feynman gauge ($\alpha=1$) but, as is not complicated to see,
it is easy to generalize the results to an arbitrary $\alpha$-gauge
since this generalization concerns only the standard part of the propagator
obtained and has been repeatedly discussed in literature 
(see, e.g., Ref. \cite{Ryd85} and references therein).

Finally there are four important remarks.

Firstly, the fact is that the notion of propagator makes no sense for general 
nonlinear equations such as the Yang-Mills ones (2.12). If restricting, however,
to the SU(3)-fields taking the values in the Cartan subalgebra of the 
SU(3)-Lee algebra (see Appendix $B$), i.e. in the subalgebra generated by the 
matrices $\lambda_3$ and $\lambda_8$, then the equations (2.12) (at $J=0$) 
become linear since the right-hand side of (2.12) is equal identically to zero 
for such field configurations (and for those gauge equivalent to the latter). 
The confining solutions (2.13)--(2.14) are just of 
the mentioned class. As a consequence, our modification of propagator holds 
true just in the latter set of SU(3)-Yang-Mills fields but the standard
gluon propagator is tacitly supposed to correspond to the given class as well  
because it conforms to the Coulomb-like part of the solutions (2.13)--(2.14).

 Secondly, one should say a few words concerning the nonrelativistic confining 
potentials often used, for example, in quarkonium theory (see, e.g., 
Ref. \cite{GM}). The confining potential between quarks here
is usually modelled in the form $a/r+br$ with some constants $a$ and $b$.
It is clear, however, that from the QCD point of view the interaction between
quarks should be described by the whole SU(3)-field $A_\mu=A^a_\mu T_a$,
genuinely relativistic object, the nonrelativistic potential being only some
component of $A^a_t$ surviving in the nonrelativistic limit when the light 
velocity $c\to\infty$.

As has been mentioned in Refs. \cite{{Gon03},{GC03}} (see also Section 8), 
however, the U(1)- or SU(3)-field of form $A^a_t=Br^\gamma$,
where $B$ is a constant, may be solution of the Maxwell and Yang-Mills 
equations (2.3), (2.12) (at $J=0$) only at $\gamma=-1$, i. e. in the 
Coulomb-like case. 

As a result, the potentials employed in nonrelativistic approaches do 
not obey the Maxwell or Yang-Mills equations.
The latter ones are essentially relativistic and, as we have seen, the
components linear in $r$ of the whole $A_\mu$ are different from $A_t$ and
related with (colour) magnetic field vanishing in the nonrelativistic limit.
That is why the nonrelativistic confining potentials cannot be assumed as a 
basis when deriving the modification of gluon propagator under consideration.

Thirdly, it should be emphasized that 
the standard gluon propagator (as well as the corresponding photon one) is by 
itself essentially {\em nonperturbative} object [a fundamental 
solution (a Green function) of d'Alembert-like system (3.2)] that cannot be 
calculable by any perturbative techniques. Another matter that standard 
propagator is used in QCD {\em perturbation} theory for calculating quantum 
corrections, including for the propagator itself. 

From mathematical point of view we should choose the Green function for the 
corresponding equations [SU(3)-Yang-Mills ones of (2.12)] which takes into 
account the necessary boundary conditions -- linear confinement which is 
essentially {\em nonperturbative} phenomenon.  The standard 
(nonperturbative) gluon propagator does not satisfy this condition. That is  
why we should modify the mentioned propagator in a {\em nonperturbative} way 
which can be done only on the basis of the corresponding exact 
{\em nonperturbative} solutions of the SU(3)-Yang-Mills equations. 
The resulting propagator in (4.1) is essentially 
{\em nonperturbative} one in each summand. 

Fourthly, the structure of the obtained propagator of (3.20) shows that it has 
rather strong infrared singularities at $k\to0$. Physical meaning of this is that 
quarks mainly emit and interchange the soft gluons (i.e., those with small $k$), 
so that gluon concentrations in the confining SU(3)-gluonic field are much 
greater than the estimates given in Section 9 (see also Ref. \cite{Gon04}) 
because the latter are the estimates for maximal possible gluon frequencies, 
i.e. for maximal possible gluon impulses (under the concrete situation of 
charmonium states). It is also clear that just magnetic 
part of the propagator (3.20) is responsible for larger portion of gluon 
concentrations since it has stronger infrared singularities than the electric 
part.

\subsection{A Scenario for Linear Confinement}
 The above results and those of 
Refs. \cite{{Gon01},{Gon03},{GC03},{Gon04}} suggest the following mechanism of 
confinement to occur within the framework of QCD (at any rate, for mesons
and quarkonia). The gluon exchange between quarks is realized by means of
the propagator described above. At small distances one may neglect 
additional addenda of order $|{\bf k}|^3$ in the propagator and
we obtain the standard gluon one used in the perturbative QCD and, as a
result, asymptotic freeedom. At large distances the mentioned gluon exchange 
leads to the confining SU(3)-field of form (2.13)--(2.14) which may be 
considered classically (the gluon concentration becomes huge and gluons form 
the boson condensate -- a classical field) and is 
a {\em nonperturbative} solution of the SU(3)-Yang-Mills 
equations. Under the circumstances mesons are the 
{\em relativistic bound states}  
described by the corresponding wave functions -- {\em nonperturbative} modulo 
square integrable solutions of the Dirac equation in this 
confining SU(3)-field \cite{{Gon01},{Gon03},{GC03},{Gon04}}. For each meson 
there exists its own 
set of real constants $a_j, A_j, b_j, B_j$ parametrizing the confining gluon
field (2.13)--(2.14) (the mentioned gluon condensate) and the corresponding 
wave functions while the latter ones also depend on $\mu_0$, the reduced
mass of the current masses of quarks forming 
meson \cite{{Gon01},{Gon03},{GC03},{Gon04}}. It is clear that constants 
$a_j, A_j, b_j, B_j,\mu_0$
should be extracted from experimental data and such a program has been just 
realized in Refs. \cite{{Gon03},{GC03},{Gon04}} for quarkonia (see also 
Section 9).

 Finally it should be emphasized that the reason why Nature chose the 
somewhat different propagator for gluons than the one for photons remains 
obscure within the framework of our considerations. To our mind, however, the 
questions of such a kind may be considered only from the cosmological positions, 
so that, for example, the problems concerning the number of quark flavours in 
QCD or a mechanism of breaking chiral symmetry in QCD can be scarcely resolved
within the QCD framework in a single-valued way and a cosmological 
approach might be quite useful \cite{GonQCD}.

\section{Confining Solutions of the SU($N$)-Yang-Mills Equations}
\subsection{Motivation}
The previous section led us to the following problem: how to describe 
possible relativistic bound states in the confining SU(3)-Yang-Mills fields? 
The sought description should be obviously based on the QCD-lagrangian. 
Let us write down this lagrangian (for one flavour and $N$ quark colours) in 
arbitrary curvilinear (local) coordinates in Minkowski spacetime 
$${\cal L}=\overline{\Psi}{\cal D}\Psi -\mu_0\overline{\Psi}\Psi 
-\frac{1}{4}(g^{\mu\alpha}g^{\nu\beta}-g^{\mu\beta}g^{\nu\alpha})
F^a_{\mu\nu}F^a_{\alpha\beta}, \,\mu<\nu,\,\alpha<\beta
\eqno(5.1)$$
where, if denoting
$S(M)$ and $\xi$, respectively, the standard spinor bundle
and $N$-dimensional vector one [equipped with a SU($N$)-connection with 
the corresponding connection and curvature matrices 
$A=A_\mu dx^\mu=A^a_\mu T_adx^\mu$, 
$F=dA+gA\wedge A=F^a_{\mu\nu}T_adx^\mu\wedge dx^\nu$, 
see Appendix $B$] over Minkowski spacetime, we can
construct tensorial product $\Xi=S(M)\otimes\xi$. It is clear that $\Psi$ is 
just a section of the latter bundle, i. e. 
$\Psi$ can be chosen in the form
$\Psi=(\Psi_1,..., \Psi_N)$
with the four-dimensional Dirac spinors $\psi_j$ representing the $j$-th colour
component while $\overline{\Psi}=\Psi^{\dag}(\gamma^0\otimes I_N)$ is 
the adjont spinor, ($\dag$) stands for hermitian conjugation, $I_N$ is the 
unit matrix $N\times N$, $\otimes$ means tensorial product of matrices,  
$\mu_0$ is a mass parameter, ${\cal D}$ is the Dirac operator
with coefficients in $\xi$ (see below). At last, we take 
the condition ${\rm Tr}(T_aT_b)=K\delta_{ab}$ with some real $K$ so that the 
third addendum in (5.1) has the form $G(F,F)/(4K)$ (see Appendix $A$), where 
coefficient $1/(4K)$ is chosen from physical considerations.

Of course, the most physically important case is that of group SU(3), i.e.,
three colors of quarks, however, it makes
sense to have analysed the general SU($N$)-case with arbitrary $N$ as well
because during a long time there is a firm belief in that considering the
limit $N\to\infty$ to be rather (or even extremely) important for
understanding of the real 4D QCD with three colors. The spectrum of
speculations on this topic ranges from phenomenological and lattice
approaches (for more details see, e. g., Refs. \cite{Ya} and references therein)
to the exotic scenarios based on $M$-theory and strings
(see, e. g., Ref. \cite{Ah} and references therein).

From general considerations  the explicit form of
the operator ${\cal D}$ in local coordinates $x^\mu$ on Minkowski spacetime 
can be written as follows
$${\cal D}=i(\gamma^e\otimes I_N)E_e^\mu\left(\partial_\mu\otimes I_N
-\frac{1}{2}\omega_{\mu ab}\gamma^a\gamma^b\otimes I_N-igA_\mu\right),
\>a < b ,\>\eqno(5.2)$$
where $g$ is a gauge coupling constant, 
the forms\ $\omega_{ab}=\omega_{\mu ab}dx^\mu$ obey 
the Cartan structure equations
$de^a=\omega^a_{\ b}\wedge e^b$ with exterior derivative $d$, while the
orthonormal basis $e^a=e^a_\mu dx^\mu$ in cotangent bundle and
dual basis $E_a=E^\mu_a\partial_\mu$ in tangent bundle are connected by the
relations $e^a(E_b)=\delta^a_b$. At last, matrices $\gamma^a$ represent
the Clifford algebra of
the corresponding quadratic form in ${\Bbb C}^{4}$. Below we shall deal only
with 4D lorentzian case (quadratic form $Q_{1,3}=x_0^2-x_1^2-x_2^2-x_3^2$).
For this we take the following choice for $\gamma^a$
$$\gamma^0=\pmatrix{1&0\cr 0&-1\cr}\,,
\gamma^b=\pmatrix{0&\sigma_b\cr-\sigma_b&0\cr}\,,
b= 1,2,3\>, \eqno(5.3)$$
where $\sigma_b$ denote the ordinary Pauli matrices (see Appendix $B$).
It should be noted that, in lorentzian case, Greek indices $\mu,\nu,...$
are raised and lowered with $g_{\mu\nu}$ of (1.1) or its inverse $g^{\mu\nu}$
and Latin indices $a,b,...$ are raised and lowered by
$\eta_{ab}=\eta^{ab}$= diag(1,-1,-1,-1) except for Latin indices connected 
with SU($N$)-Lie algebras, 
so that $e^a_\mu e^b_\nu g^{\mu\nu}=\eta^{ab}$,
$E^\mu_aE^\nu_bg_{\mu\nu}=\eta_{ab}$ and so on but $T_a=T^a$.

Under the circumstances we can obtain the following equations according to the 
standard prescription of Lagrange approach from (5.1)
$${\cal D}\Psi=\mu_0\Psi\>,\eqno(5.4)$$
$$d\ast F= g(\ast F\wedge A - A\wedge\ast F) +gJ\>,\eqno(5.5)$$
where $\ast$ means the Hodge star
operator conforming to a Minkowski metric, for instance, in the form of (1.1), 
while the source $J$ (a nonabelian SU($N$)-current) is 
$$J=j_\mu^aT_a\ast(dx^\mu)=\ast j=\ast(j_\mu^aT_adx^\mu)=
\ast(j^aT_a)\>\eqno(5.6)$$
where currents 
$$j^a=j_\mu^adx^\mu=\overline{\Psi}(\gamma_\mu\otimes I_N)T^a\Psi\,dx^\mu\>,$$
so summing over 
$a=1,...,N^2-1$ is implied in (5.1) and (5.6). 

\hbox{When using the relation (see, e. g. Refs. \cite{Fin})}
$\gamma^cE_c^\mu\omega_{\mu ab}\gamma^a\gamma^b=
\omega_{\mu ab}\gamma^\mu\gamma^a\gamma^b=
-{\rm div}(\gamma)$ with matrix 1-form $\gamma=\gamma_\mu dx^\mu$ 
[where ${\rm div}$ is defined by relation ($B$.4)] and also the 
fact that $(\gamma^\mu)^{\dag}\gamma^0=\gamma^0\gamma^\mu$, the 
Dirac equation for spinor $\overline{\Psi}$ will be  
$$i\partial_\mu\overline{\Psi}(\gamma^\mu\otimes I_N)+
\frac{i}{2}\overline{\Psi}{\rm div}(\gamma)\otimes I_N-
g\overline{\Psi}(\gamma^\mu\otimes I_N)A^a_\mu T_a=-\mu_0\overline{\Psi}
\>.\eqno(5.4')$$
Then multiplying (5.4) by $\overline{\Psi}T_a$ from left and (5.4$^\prime$) by 
$T_a\Psi$ from right and adding the obtained equations, we get 
${\rm div}(j^a)={\rm div}(j)=0$ if spinor ${\Psi}$ obeys the Dirac equation 
(5.4).

The question now is how to connect the sought relativistic bound states with 
the system (5.4)--(5.5). To understand it let us apply to the experience 
related with QED. In the latter case lagrangian looks like (5.1) with changing 
group SU($N$)$\to$U(1) so $\Psi$ will be just a four-dimensional Dirac spinor. 
Then, as is known (see, e. g. Ref. \cite{LL1}), when passing on to the 
nonrelativistic limit the Dirac equation (5.4) converts into the Pauli equation and 
further, if neglecting the particle spin, into the Schr{\"o}dinger 
equation, parameter $\mu_0$ becoming the reduced mass of two-body system. 
The modulo square integrable solutions of the Schr{\"o}dinger equation just 
describe bound states of a particle with mass $\mu_0$, or, that is equivalent, 
of the corresponding two-body system. Historically, however, everything was 
just vice versa. At first there appeared the Schr{\"o}dinger equation, then 
the Pauli and Dirac ones and only then the QED lagrangian. In its turn, 
possibility of writing two-body Schr{\"o}dinger equation on the whole owed 
to the fact that the corresponding two-body problem in classical 
nonrelativistic (newtonian) mechanics was well posed and actually quantizing 
the latter gave two-body Schr{\"o}dinger equation. Another matter was Dirac 
equation. Up to now nobody can say what two-body problem in classical 
relativistic (einsteinian) mechanics could correspond to Dirac equation. The 
fact is that the two-body problem in classical relativistic mechanics has 
so far no single-valued statement. Conventionally, therefore, Dirac 
equation in QED is treated as the relativistic wave equation describing one 
particle with spin one half in an external electromagnetic field. 

There is, however, one important exclusion -- the hydrogen atom. When solving 
the Dirac equation here one considers mass parameter $\mu_0$ to be equal to the 
electron mass and one gets the so-called Sommerfeld formula for hydrogen atom 
levels which passes on to the standard Schr{\"o}dinger formula for hydrogen atom 
spectrum in nonrelativistic limit (for more details see, e. g. 
Ref. \cite{LL1} and also Subsection 6.2 ). But in the Schr{\"o}dinger formula 
mass parameter $\mu_0$ is equal to the reduced mass of electron and proton. As 
a consequence, 
it is tacitly supposed that in Dirac equation the mass parameter should be 
equal to the same reduced mass of electron and proton as in Schr{\"o}dinger 
equation. Just the mentioned reduced mass is approximately equal to that of 
electron but, exactly speaking, it is not the case. We remind that for the 
problem under discussion (hydrogen atom) the external field is the Coulomb 
electric one between electron and proton, essentially nonrelativistic object 
in the sense that it does not vanish in nonrelativistic limit at $c\to\infty$.
If now to place hydrogen atom in a magnetic field then obviously spectrum of 
bound states will also depend on parameters decribing the magnetic field. 
The latter, however, is essentially relativistic object and vanishes at 
$c\to\infty$ because, as is well known, in the world with $c=\infty$ there 
exist no magnetic fields (see any elementary textbook on physics, e. g. 
Ref. \cite{Sav82}). But it is clear that spectrum should as before depend of $\mu_0$ 
as well and we can see that $\mu_0$ is the same reduced mass as before 
since in nonrelativistic limit we again should come to the hydrogen atom 
spectrum with the reduced mass. So we can draw the conclusion that if
an electromagnetic field is a combination of electric Coulomb field between two 
charged particles and some magnetic field (which may be generated by the 
particles themselves) then there are certain grounds to consider the given 
(quantum) two-body problem to be equivalent to the one of motion for one 
particle with usual reduced mass in the mentioned electromagnetic field. As a 
result, we can use the Dirac equation for finding possible relativistic 
bound states for such a particle implying that this is really some description 
of the corresponding two-body problem. 

Actually in QED the situaion is just as the described one but magnetic field 
is usually weak and one may restrict 
oneself to some corrections from this field to the nonrelativistic Coulomb 
spectrum (e.g., in the Zeeman effect). If the magnetic field is strong then 
one should solve just Dirac 
equation in a nonperturbative way (see, e.g. Ref. \cite{ST}). The latter 
situation seems to be natural in QCD where the corresponding magnetic (colour)  
field should be very strong 
because just it provides linear confinement of quarks as we shall see below 
(see also Section 9).

At last, we should make an important point that in QED the mentioned 
electromagnetic field is by definition always a solution of the Maxwell 
equations so 
within QCD we should require the confining SU(3)-field to be a solution of 
Yang-Mills equations. Consequently, returning to the system (5.4)--(5.5), we 
can suggest to decribe relativistic bound states of two quarks (mesons) in QCD 
by the compatible solutions of the given system. To be more precise, the meson 
wave functions should be the {\em nonperturbative} modulo square integrable 
solutions of Dirac equation (5.4) (with the above reduced mass $\mu_0$) in the 
confining SU(3)-Yang-Mills field being a {\em nonperturbative} solution of 
(5.5). In general case, however, the analysis of (5.5) is difficult because of 
availability of the nonabelian current $J$ of (5.6) in the right-hand side of 
(5.5) but we may use the circumstance that the corresponding modulo square 
integrable solutions of Dirac equation (5.4) might consist from the components 
of form $\Psi_j\sim r^{\alpha_j}e^{-\beta_j r}$ with some $\alpha_j>0$, 
$\beta_j>0$ 
which entails all the components of the current $J$ to be modulo $<< 1$ at each 
point of Minkowski space. The latter will allow us to put $J\approx0$ and we 
shall come to the problem of finding the confining solutions for the Yang-Mills 
equations of (5.5) with $J=0$ (according to Section 1 such solutions are  
supposed to be spherically symmetric and to contain only the components of the 
SU($N$)-field which are Coulomb-like or linear in $r$) and after inserting the 
found solutions into Dirac equation (5.4) we should require the corresponding 
solutions of Dirac equation to have the above necessary behaviour. Under 
the circumstances the problem becomes self-consistent and can be analyzable.

It is clear that all the above considerations can be justified only by  
comparison with experimental data but now we obtained some intelligible 
programme of further activity. So let us pass on to its realization.

\subsection{Role of Diagonal Gauge}
 Let us in detail write out $A_\mu=A^a_\mu T_a$ of the Dirac operator 
from (5.2) employing the SU($N$)-Lie algebra realizations from Appendix $B$. 
We obtain at $N= 2, 3, 4$ respectively
$$A^a_\mu\sigma_a=
\pmatrix
{A^3_\mu &A^1_\mu-iA^2_\mu\cr
A^1_\mu+iA^2_\mu&-A^3_\mu\cr}\>,$$
$$A^a_\mu\lambda_a=
\pmatrix
{A^3_\mu+\frac{1}{\sqrt{3}}A^8_\mu&A^1_\mu-iA^2_\mu&A^4_\mu-iA^5_\mu\cr
A^1_\mu+iA^2_\mu&-A^3_\mu+\frac{1}{\sqrt{3}}A^8_\mu&A^6_\mu-iA^7_\mu\cr
A^4_\mu+iA^5_\mu&A^6_\mu+iA^7_\mu&-\frac{2}{\sqrt{3}}A^8_\mu\cr}\>,$$
$$A^a_\mu T_a=$$
$$\pmatrix
{A^3_\mu+A^6_\mu+A^{15}_\mu & z_1 & z_2 & z_3 \cr
z_1^\ast & -A^3_\mu+A^6_\mu-A^{15}_\mu & z_4 & z_5 \cr 
z_2^\ast & z_4^\ast & A^3_\mu-A^6_\mu-A^{15}_\mu & z_6 \cr
z_3^\ast & z_5^\ast & z_6^\ast & -A^3_\mu-A^6_\mu+A^{15}_\mu \cr} 
\> \eqno(5.7)$$
with $z_1=A^1_\mu+A^9_\mu-i(A^2_\mu+A^{12}_\mu)$, 
$z_2=A^4_\mu+A^{13}_\mu-i(A^5_\mu+A^{14}_\mu)$, 
$z_3=A^7_\mu-A^{11}_\mu-i(A^8_\mu+A^{10}_\mu)$,  
$z_4=A^7_\mu+A^{11}_\mu-i(A^8_\mu-A^{10}_\mu)$, 
$z_5=A^4_\mu-A^{13}_\mu-i(A^5_\mu-A^{14}_\mu)$, 
$z_6=A^1_\mu-A^{9}_\mu-i(A^2_\mu-A^{12}_\mu)$, where (*) signifies complex 
conjugation. Then it is natural to put 
$A^a_\mu=0$ with $a=1,2$ at $N=2$, $A^a_\mu=0$ with $a=1,2,4,5,6,7$ at $N=3$, 
$A^a_\mu=0$ with $a=1,2,4,5,7,8,9,10,11,12,13,14$ at $N=4$ so long as 
the Dirac equation (5.4) in such a gauge takes the simplest form. We further 
call this gauge {\em diagonal} one. Really Dirac equation (5.4) in diagonal 
gauge splits into the system of Dirac equations for components $\Psi_j$. 
Namely, at $N=2$ the system is 
$$i\gamma^eE^\mu_e\left[\partial_\mu
-\frac{1}{2}\omega_{\mu ab}\gamma^a\gamma^b-igA^3_\mu\right]\Psi_1
=\mu_0\Psi_1 \>,$$
$$i\gamma^eE^\mu_e\left[\partial_\mu
-\frac{1}{2}\omega_{\mu ab}\gamma^a\gamma^b+igA^3_\mu
\right]\Psi_2 =\mu_0\Psi_2 \>,\eqno(5.8)$$
while at $N=3$ it is 
$$i\gamma^eE^\mu_e\left[\partial_\mu
-\frac{1}{2}\omega_{\mu ab}\gamma^a\gamma^b-ig\left(A^3_\mu+
\frac{1}{\sqrt{3}}A^8_\mu\right)\right]\Psi_1 =\mu_0\Psi_1 \>,$$
$$i\gamma^eE^\mu_e\left[\partial_\mu
-\frac{1}{2}\omega_{\mu ab}\gamma^a\gamma^b-ig\left(-A^3_\mu+
\frac{1}{\sqrt{3}}A^8_\mu\right)\right]\Psi_2 =\mu_0\Psi_2 \>,$$
$$i\gamma^eE^\mu_e\left[\partial_\mu
-\frac{1}{2}\omega_{\mu ab}\gamma^a\gamma^b-
ig\left(-\frac{2}{\sqrt{3}}A^8_\mu\right)\right]\Psi_3=\mu_0\Psi_3
\>,\eqno(5.9)$$
and, at last, at $N=4$ the corresponding system is 
$$i\gamma^eE^\mu_e\left[\partial_\mu
-\frac{1}{2}\omega_{\mu ab}\gamma^a\gamma^b-
ig\left(A^3_\mu+A^6_\mu+A^{15}_\mu\right)\right]\Psi_1 =\mu_0\Psi_1 \>,$$
$$i\gamma^eE^\mu_e\left[\partial_\mu
-\frac{1}{2}\omega_{\mu ab}\gamma^a\gamma^b-
ig\left(-A^3_\mu+A^6_\mu-A^{15}_\mu\right)\right]\Psi_2 =\mu_0\Psi_2 \>,$$
$$i\gamma^eE^\mu_e\left[\partial_\mu
-\frac{1}{2}\omega_{\mu ab}\gamma^a\gamma^b-
ig\left(A^3_\mu-A^6_\mu-A^{15}_\mu\right)\right]\Psi_3=\mu_0\Psi_3\>,$$
$$i\gamma^eE^\mu_e\left[\partial_\mu
-\frac{1}{2}\omega_{\mu ab}\gamma^a\gamma^b-
ig\left(-A^3_\mu-A^6_\mu+A^{15}_\mu\right)\right]\Psi_4=\mu_0\Psi_4
\>.\eqno(5.10)$$

It is clear that in diagonal gauge the 
SU$(N)$-Yang-Mills fields are described by matrices $A_\mu=A^a_\mu T_a$ taking 
their values in the Cartan subalgebra of the conforming SU($N$)-Lie algebra 
(see Appendix $B$). 

Due to that $dx^\mu\wedge dx^\nu=-dx^\nu\wedge dx^\mu$, we have 
$A\wedge A=A^a_\mu A^b_\nu[T_a,T_b]dx^\mu\wedge~dx^\nu$, $\mu<\nu$.
Consequently for the SU$(N)$-Yang-Mills with values in the Cartan subalgebra 
$A\wedge A=0$ since commutator $[T_a,T_b]$ for diagonal matrices 
is always equal 
to zero and the Cartan subalgebras of SU$(N)$-groups just consist from 
diagonal matrices (see Appendix $B$). Accordingly the curvature matrix (field 
strength) $F=dA+gA\wedge A=dA$ while the right-hand 
side of the Yang-Mills equations (5.5) (at $J=0$) is identically equal to zero 
since matrix $\ast F$ is also diagonal and then 
$\ast F\wedge A = A\wedge\ast F$. 
This fact strongly simplifies the task of searching for confining solutions 
because the equations (5.5) convert into 
$$d\ast F=0\>. \eqno(5.11) $$  
It should be emphasized that such a simplification is dictated by the wish to
obtain the simplest form for the Dirac equation (5.4). Clearly, all the results 
obtained in diagonal gauge will hold true in any gauge connected with the 
diagonal one by some gauge transfomation due to the fact of gauge invariance 
of the Yang-Mills equations (5.5). It turns out, however, that the confining 
solutions obtained in this way has the property of uniqueness in a certain 
sense which will be discussed in Section 8 and now let us pass on to finding 
confining solutions.

\subsection{U(1)-case}
  We have already discussed this case in Section 2, where the corresponding 
confining solution was found in the form (2.6). But it could seem that 
when searching for the solution the ansatz used was not the most general one. 
Really, we took the ansatz in the form $A=A_t(r)dt+A_\varphi(r)d\varphi$. It 
seems that the most general form is $A=A_t(r)dt+A_r(r)dr+
A_\vartheta(r)d\vartheta+A_\varphi(r)d\varphi$. Let us discuss it in more 
details. 

For the latter ansatz we have $F=dA=-\partial_rA_tdt\wedge dr+
\partial_rA_\vartheta dr\wedge d\vartheta+
\partial_rA_\varphi dr\wedge d\varphi$ 
for an arbitrary $A_r(r)$. Then, according to ($A.6$), we obtain 
$$\ast F= (r^2\sin{\vartheta})\partial_rA_td\vartheta\wedge d\varphi +
\sin{\vartheta}\partial_rA_\vartheta dt\wedge d\varphi-
\frac{1}{\sin{\vartheta}}\partial_rA_\varphi dt\wedge d\vartheta \eqno(5.12)$$ 
which entails 
$$d\ast F= 
\sin{\vartheta}\partial_r(r^2\partial_rA_t)dr\wedge d\vartheta\wedge d\varphi-
\sin{\vartheta}\partial_r^2A_\vartheta dt\wedge dr\wedge d\varphi-$$
$$\cos{\vartheta}\partial_rA_\vartheta dt\wedge d\vartheta\wedge d\varphi+
\frac{1}{\sin{\vartheta}}\partial_r^2A_\varphi dt\wedge dr\wedge d\vartheta
=0\>,\eqno(5.13)$$
wherefrom one can conclude that
$$\partial_r(r^2\partial_rA_t)=0,\>\partial^2_rA_\varphi=0\>,\eqno(5.14)$$
$$\partial^2_rA_\vartheta=\partial_rA_\vartheta=0\>.\eqno(5.15)$$
This yields the solutions (2.6) while we draw the conclusion that 
$A_\vartheta=C_1$ with some constant $C_1$. But then the Lorentz 
condition (2.2) for the given ansatz entails 
$$\sin{\vartheta}\partial_r(r^2A_r)+
\partial_\vartheta(\sin{\vartheta}A_\vartheta)=0, $$
or 
$$\partial_r(r^2A_r)+
\cot{\vartheta}A_\vartheta=0, \eqno(5.16)$$
which entails $A_r=-A_\vartheta\cot{\vartheta}/r+C_2/r^2$ with a 
constant $C_2$. But the confining solutions, as we accept in the given paper, 
should be spherically symmetric and contain only the components which are 
Coulomb-like or linear in $r$, so one should put $C_1=C_2=0$. Consequently, 
the ansatz $A=A_t(r)dt+A_\varphi(r)d\varphi$ is most general 
and we can consider $A_r=A_\vartheta=0$ without loss of generality.

Let us describe one class of the confining nonspherically symmetric solutions 
of the Maxwell equations (2.3) (at $J=0$) that can be obtained with the aid
of the ansatz $A=A_t(r)dt+A_\varphi(r,\vartheta)d\varphi$, i. e. now we
consider the component $A_\varphi$ depending also on $\vartheta$. 
It is evident that the Lorentz condition (2.2) is automatically fulfilled 
for the given ansatz. We shall have $F=dA= 
-\partial_rA_tdt\wedge dr+\partial_rA_\varphi dr\wedge d\varphi
+\partial_\vartheta A_\varphi d\vartheta\wedge d\varphi$
and $\ast F=r^2\sin\vartheta\partial_rA_td\vartheta\wedge d\varphi-
\frac{1}{\sin\vartheta}\partial_rA_\varphi dt\wedge d\vartheta
+\frac{1}{r^2\sin\vartheta}\partial_\vartheta A_\varphi dt\wedge dr$. Then
Eq. (2.3) (at $J=0$) entails
$$\partial_r(r^2\partial_rA_t)=0,\>\eqno(5.17)$$
$$r^2\partial^2_rA_\varphi
+\sin\vartheta\partial_\vartheta\left(\frac{1}{\sin\vartheta}
\partial_\vartheta A_\varphi\right)=0\>.\eqno(5.18)$$
We shall not here discuss the general form of the solution for Eq. (5.18) and 
only write out the possible solution of (5.17)--(5.18) which is useful to us 
in the present paper in the form slightly modifying (2.6)
$$ A_t =\frac{a}{r}+A \>, A_\varphi=br+B -K\cos{\vartheta}\>\eqno(5.19)$$
with some constants $a, b, A, B, K$ parametrizing solution.

\subsection{$N=2$}
Remarks done in previous subsection about the Lorentz condition will hold 
true for any group SU($N$) (see Subsection 8.1), so at $N=2$ we put 
$A^{3}_{r,\vartheta}=0$. After 
this we search
for the solution of ($B.3$) (at $J=0$) in the form 
$A=A_t(r)dt+A_\varphi(r)d\varphi$
with $A_{t,\varphi}=A^3_{t,\varphi}\sigma_3$. Along the above lines 
it is then easy to come to the system 
$$\partial_r(r^2\partial_rA_t)=0,\>\partial^2_rA_\varphi=0\>,\eqno(5.20)$$
and we write down the solutions of (5.20) needed to insert into (5.8) 
$$ A^3_t =-\frac{a}{r}+A \>,\eqno(5.21)$$
$$ A^3_\varphi =br+B \>\eqno(5.22)$$
with some constants $a, A, b, B$ parametrizing solutions.

Class of the confining nonspherically symmetric solutions of ($B.3$) can be 
obtained with the aid of the ansatz 
$A=A_t(r)dt+A_\varphi(r,\vartheta)d\varphi$.
We shall have 
$$\partial_r(r^2\partial_rA_t)=0,\>\eqno(5.23)$$
$$r^2\partial^2_rA_\varphi
+\sin\vartheta\partial_\vartheta\left(\frac{1}{\sin\vartheta}
\partial_\vartheta A_\varphi\right)=0\>.\eqno(5.24)$$
It is clear that Eq. (5.23) gives the same solution of (5.21) while 
one possible solution of (5.24) useful to us in the present
paper is
$$ A^3_\varphi=-K\cos\vartheta+br+B \>\eqno(5.25)$$
with some real constants $K, b, B$ parametrizing solution.

\subsection{$N=3$}
 In the given case we put $A^{3,8}_{r,\vartheta}=0$. After this the ansatz
in the form $A=A_t(r)dt+A_\varphi(r)d\varphi$
with $A_{t,\varphi}=A^3_{t,\varphi}\lambda_3+A^8_{t,\varphi}\lambda_8$ 
yields (at $J=0$) the solutions (2.13). The corresponding class of the 
confining nonspherically symmetric solutions of ($B.3$) (at $J=0$)
can be obtained with the aid of the ansatz 
$A=A_t(r)dt+A_\varphi(r,\vartheta)d\varphi$ which gives the same component 
$A_t$ as in (2.13) while
$$ A^3_\varphi+\frac{1}{\sqrt{3}}A^8_\varphi =-K_1\cos\vartheta+b_1r+B_1 \>,$$
$$ -A^3_\varphi+\frac{1}{\sqrt{3}}A^8_\varphi=-K_2\cos\vartheta+b_2r+B_2\>,$$
$$-\frac{2}{\sqrt{3}}A^8_\varphi=(K_1+K_2)\cos\vartheta-(b_1+b_2)r-(B_1+B_2)
\> \eqno(5.26)$$
with some real constants $K_j, b_j, B_j$ parametrizing solution.

\subsection{$N=4$}
We put $A^{3,6,15}_{r,\vartheta}=0$ and at $J=0$ the ansatz 
$A=A_t(r)dt+A_\varphi(r)d\varphi$
with $A_{t,\varphi}=A^3_{t,\varphi}T_3+A^6_{t,\varphi}T_8+
A^{15}_{t,\varphi}T_{15}$ gives rise to the solutions of ($B.3$) in the 
form 
 $$ A^3_t+A^6_t+A^{15}_t =-\frac{a_1}{r}+A_1 \>,
 -A^3_t+A^6_t-A^{15}_t=-\frac{a_2}{r}+A_2\>,$$
$$A^3_t-A^6_t-A^{15}_t=-\frac{a_3}{r}+A_3\>,
-A^3_t-A^6_t+A^{15}_t=\frac{a_1+a_2+a_3}{r}-(A_1+A_2+A_3)\>,$$
 $$ A^3_\varphi+A^6_\varphi+A^{15}_\varphi =b_1r+B_1 \>,
 -A^3_\varphi+A^6_\varphi-A^{15}_\varphi=b_2r+B_2\>,$$
$$A^3_\varphi-A^6_\varphi-A^{15}_\varphi=b_3r+B_3\>,
-A^3_\varphi-A^6_\varphi+A^{15}_\varphi=-(b_1+b_2+b_3)r-(B_1+B_2+B_3)\>,
\eqno(5.27)$$
where real constants $a_j, A_j, b_j, B_j$
parametrize the solutions. The corresponding class of the confining 
nonspherically symmetric solutions of ($B.3$) 
(at $J=0$) can be obtained with the aid of the ansatz 
$A=A_t(r)dt+A_\varphi(r,\vartheta)d\varphi$ which gives the same component
$A_t$ as in (5.27) while
$$ A^3_\varphi+A^6_\varphi+A^{15}_\varphi =b_1r+B_1-K_1\cos{\vartheta} \>,
 -A^3_\varphi+A^6_\varphi-A^{15}_\varphi=b_2r+B_2-K_2\cos{\vartheta}\>,$$
$$A^3_\varphi-A^6_\varphi-A^{15}_\varphi=b_3r+B_3-K_3\cos{\vartheta}\>,$$
$$-A^3_\varphi-A^6_\varphi+A^{15}_\varphi=-(b_1+b_2+b_3)r-(B_1+B_2+B_3)+
(K_1+K_2+K_3)\cos{\vartheta}\>
\eqno(5.28)$$
with real constants $K_j, b_j, B_j$.

\section{Spectrum of Bound States in the Coulomb-Like Case}
The question now is how to find the modulo square integrable solutions of Dirac 
equation (5.4) when inserting the confining solutions described in 
previous Section into it. We shall need some results about spectrum of the 
euclidean Dirac operator on the unit two-dimensional sphere ${\Bbb S}^2$ in 
the form obtained in Refs. \cite{Gon99}. 
\subsection{Results from the Black Hole Theory about 
Eigenspinors of the (Twisted) Euclidean Dirac Operator 
on ${\Bbb S}^2$} 
When separating variables in (5.4) (see next Subsection) there naturally 
arises the euclidean Dirac operator ${\cal D}_0$ on the unit two-dimensional 
sphere ${\Bbb S}^2$ and we should know its eigenvalues with the corresponding 
eigenspinors. Such a problem also arises in the black hole theory while 
describing the so-called twisted spinors on Schwarzschild and 
Reissner-Nordstr\"om black holes and it was analysed in 
Refs. \cite{Gon99}, so we can use the results obtained therein\linebreak for our aims.  
Let us adduce the necessary relations. 

Let us consider $2k$-dimensional (pseudo)riemannian manifold $M$ for which 
$H^1(M,{\Bbb Z}_2)$, the first cohomology group with coefficients in 
${\Bbb Z}_2$, is equal to zero while $H^2(M,{\Bbb Z})$, the second cohomology 
group with coefficients in ${\Bbb Z}$, is equal to ${\Bbb Z}$. Then standard 
topological results \cite{{81},{89},{Bes87}} say to us that over $M$ there 
exists the only so-called Spin-structure whereas there is countable number of 
complex line bundles over $M$. As a consequence, each complex line bundle can 
be characterized by an integer $n$ which in what follows will be called its 
Chern number. Under this situation, if denoting $S(M)$ the only standard 
spinor bundle over $M$ and $\xi$ the complex line bundle with Chern number $n$, 
we can construct tensorial product $S(M)\otimes\xi$. Under the circumstances 
we obtain the {\it twisted Dirac operator}
${\cal D}_n: S(M)\otimes\xi\to S(M)\otimes\xi$, so the eigenvalue equation for
corresponding spinors $\Phi$ as sections of the bundle
$S(M)\otimes\xi$ may look as follows
$${\cal D}_n\Phi=\lambda\Phi,\>\eqno(6.1)$$
and we can call (standard) spinors corresponding to $n=0$ (trivial complex
line bundle $\xi$) {\it untwisted} while the rest of the spinors with $n\ne0$
should be referred to as {\it twisted}.

 From general considerations \cite{{81},{89},{Bes87}} the explicit form of
the operator ${\cal D}_n$ in local coordinates $x^\mu$ on a $2k$-dimensional
(pseudo)riemannian manifold can be written as follows
$${\cal D}_n=i\gamma^\mu\nabla_\mu\equiv i\gamma^cE_c^\mu(\partial_\mu-
\frac{1}{2}\omega_{\mu ab}\gamma^a\gamma^b-ieA_\mu),\>a < b ,\>\eqno(6.2)$$
where $A=A_\mu dx^\mu$ is a connection in the bundle $\xi$ and the forms
$\omega_{ab}=\omega_{\mu ab}dx^\mu$ obey the Cartan structure equations
$de^a=\omega^a_{\ b}\wedge e^b$ with exterior derivative $d$, while the
orthonormal basis $e^a=e^a_\mu dx^\mu$ in cotangent bundle and
dual basis $E_a=E^\mu_a\partial_\mu$ in tangent bundle are connected by the
relations $e^a(E_b)=\delta^a_b$. At last, matrices $\gamma^a$ represent
the Clifford algebra of the corresponding quadratic form in ${\Bbb C}^{2^k}$. 
Below we shall deal only with 2D euclidean case of the unit sphere 
${\Bbb S}^2$ ($k=1$, quadratic form $Q_2=x_0^2+x_1^2$).

As for the connection $A_\mu$ in bundle $\xi$ then the suitable one can be 
found, for example, in Refs. \cite{Gon678} and is
$$A= A_\mu dx^\mu=-\frac{n}{e}\cos\vartheta d\varphi\>.
\eqno(6.3)$$
Under the circumstances, as was shown in Refs. \cite{Gon678},
integrating $F=dA$ over the unit sphere ${\Bbb S}^2$ gives rise to the Dirac 
charge quantization condition
$$\int_{{\Bbb S}^2} F=4\pi\frac{n}{e}=4\pi q \eqno(6.4)$$
with magnetic charge $q$,
so we can identify the coupling constant $e$ with electric charge. 

As was discussed in Refs. \cite{Gon99}, the natural form of ${\cal D}_n$ in 
local coordinates $\vartheta, \varphi$ on the unit sphere ${\Bbb S}^2$ looks 
as follows
$${\cal D}_n=-i\sigma_1\left[
i\sigma_2\partial_\vartheta+i\sigma_3\frac{1}{\sin{\vartheta}}
\left(\partial_\varphi-\frac{1}{2}\sigma_2\sigma_3\cos{\vartheta}+
in\cos{\vartheta}\right)\right]=$$
$$\sigma_1\sigma_2\partial_\vartheta+\frac{1}{\sin\vartheta}
\sigma_1\sigma_3\partial_\varphi- \frac{\cot\vartheta}{2}
\sigma_1\sigma_2+in\sigma_1\sigma_3\cot\vartheta
         \eqno(6.5)$$
with the Pauli matrix $\sigma_j$ (see Appendix $B$), 
so that $\sigma_1{\cal D}_n=-{\cal D}_n\sigma_1$.
As is not complicated to see, the operator ${\cal D}_n$ has the form (6.2) 
with $\gamma^0=-i\sigma_1\sigma_2$, $\gamma^1=-i\sigma_1\sigma_3$,
$e^0=d\vartheta$, $e^1=\sin{\vartheta}d\varphi$,$E_0=\partial_\vartheta$, 
$E_1=\partial_\varphi/\sin{\vartheta}$, 
$\omega_{01}=\cos{\vartheta}d\varphi$,
$A_\mu dx^\mu= -\frac{n}{e}\cos{\vartheta}d\varphi$.

The equation (6.1) was explored in Refs. \cite{Gon99}.
Spectrum of $D_n$ consists of the numbers
$\lambda=\pm\sqrt{(l+1)^2-n^2}$              
with multiplicity $2(l+1)$ of each one, where $l=0,1,2,...$, $l\ge|n|$. Let us 
introduce the number $m$ such that $-l\le m\le l+1$ and the corresponding 
number $m'=m-1/2$ so $|m'|\le l+1/2$. Then the conforming eigenspinors of  
operator ${\cal D}_n$ are 
$$\Phi=\pmatrix{\Phi_1\cr\Phi_2\cr}= 
\Phi_{\mp\lambda}=\frac{C}{2}\pmatrix{P^k_{m'n-1/2}\pm P^k_{m'n+1/2}\cr
P^k_{m'n-1/2}\mp P^k_{m'n+1/2}\cr}e^{-im'\varphi}\> \eqno(6.6) $$
with the coefficient $C=\sqrt{\frac{l+1}{2\pi}}$. 
These spinors form an orthonormal basis in $L_2^2({\Bbb S}^2)$ for each $n$ 
and are subject 
to the normalization condition
$$\int_{{\Bbb S}^2}\Phi^{\dag}\Phi d\Omega=
\int\limits_0^\pi\,\int\limits_0^{2\pi}(|\Phi_{1}|^2+|\Phi_{2}|^2)
\sin\vartheta d\vartheta d\varphi=1\> , \eqno(6.7)$$
where ($\dag$) stands for hermitian conjugation.
As to functions $P^k_{m'n'}(\cos\vartheta)$ then they can be chosen by 
miscellaneous ways, for instance, as follows (see, e. g.,
Ref. \cite{Vil91})
$$P^k_{m'n'}(\cos\vartheta)=i^{-m'-n'}
\sqrt{\frac{(k-m')!(k-n')!}{(k+m')!(k+n')!}}
\left(\frac{1+\cos{\vartheta}}{1-\cos{\vartheta}}\right)^{\frac{m'+n'}{2}}\,
\times$$
$$\times\sum\limits_{j={\rm{max}}(m',n')}^k
\frac{(k+j)!i^{2j}}{(k-j)!(j-m')!(j-n')!}
\left(\frac{1-\cos{\vartheta}}{2}\right)^j \eqno(6.8)$$
with the orthogonality relation at $m',n'$ fixed
$$\int\limits_0^\pi\,{P^{*k}_{m'n'}}(\cos\vartheta)
P^{k'}_{m'n'}(\cos\vartheta)
\sin\vartheta d\vartheta={2\over2k+1}\delta_{kk'}
\>,\eqno(6.9)$$
where (*) signifies complex conjugation. It should be noted that square of 
${\cal D}_n$ is 
$${\cal D}^2_n = {\cal D}^2_0 -\frac{2in\cos{\vartheta}\partial_\varphi-n^2}
{\sin^2{\vartheta}}+in\frac{1}{\sin^2{\vartheta}}\sigma_2\sigma_3 -n^2
\eqno(6.10)$$
with 
$${\cal D}^2_0=-\Delta_{{\Bbb S}^2}+
\sigma_2\sigma_3\frac{\cos{\vartheta}}{\sin^2{\vartheta}}\partial_\varphi
+\frac{1}{4\sin^2{\vartheta}} +\frac{1}{4}\>,
\eqno(6.11)$$
while laplacian on the unit sphere is
$$\Delta_{{\Bbb S}^2}=
\frac{1}{\sin{\vartheta}}\partial_\vartheta\sin{\vartheta}\partial_\vartheta+
\frac{1}{\sin^2{\vartheta}}\partial^2_\varphi=
\partial^2_\vartheta+\cot{\vartheta}\partial_\vartheta
+\frac{1}{\sin^2{\vartheta}}\partial^2_\varphi\>,
\eqno(6.12)$$
so the relation (6.10) is a particular case of the so-called 
Weitzenb{\"o}ck-Lichnerowicz formulas (see Refs. \cite{{81},{89},{Bes87}}). 
Then from (6.1) it follows 
${\cal D}^2_n\Phi=\lambda^2\Phi$ and, when using the ansatz  
$\Phi=P(\vartheta)e^{-im'\varphi}=\pmatrix{P_1\cr P_2\cr}e^{-im'\varphi}$, 
$P_{1,2}=P_{1,2}(\vartheta)$, the equation ${\cal D}^2_n\Phi=\lambda^2\Phi$ 
turns into 
$$\left(-\partial^2_\vartheta-\cot{\vartheta}\partial_\vartheta +
\frac{m'^2+n^2+\frac{1}{4}-2m'n\cos{\vartheta}}{\sin^2{\vartheta}}+
\frac{m'\cos{\vartheta}-n}{\sin^2{\vartheta}}\sigma_1\right)P=$$
$$\left(\lambda^2+n^2-\frac{1}{4}\right)P\>,
\eqno(6.13)$$
wherefrom all the above results concerning spectrum of ${\cal D}_n$ can be 
derived \cite{Gon99}.
\subsection{U(1)-case}
We should insert the confining solutions (2.6) into Dirac equation (5.4), 
where $\Psi$ will be just a four-dimensional Dirac spinor,
and let us employ the ansatz
$$\Psi=e^{i\omega t}r^{-1}\pmatrix{F_{1}(r)\Phi(\vartheta,\varphi)\cr\
F_{2}(r)\sigma_1\Phi(\vartheta,\varphi)}\>,\eqno(6.14)$$
with a 2D spinor $\Phi=\pmatrix{\Phi_{1}\cr\Phi_{2}}$. Then, after a simple 
matrix algebra computation, we can get from (5.4) the system
$$\left[\left(\partial_r+\frac{1}{r}\right)+\frac{1}{r}{\cal D}_0-
\frac{\sigma_2}{\sin{\vartheta}}
g\left(b+\frac{B}{r}\right)\right]\frac{1}{r}F_{1}\Phi=
i(\mu_0-c)\frac{1}{r}F_{2}\Phi,$$
$$\left[\left(\partial_r+\frac{1}{r}\right)+\frac{1}{r}{\cal D}_0-
\frac{\sigma_2}{\sin{\vartheta}}
g\left(b+\frac{B}{r}\right)\right]\frac{1}{r}F_{2}\sigma_1\Phi=
-i(\mu_0+c)\frac{1}{r}F_{1}\sigma_1\Phi\eqno(6.15)$$
with $c=\omega-g(-a/r+A)$ while the euclidean Dirac operator ${\cal D}_0$ on 
the unit sphere ${\Bbb S}^2$ is given by (6.5) at $n=0$.
It is not complicated to check that at $b\ne0$, $B\ne0$ the variables $r$ and
$\vartheta$ are not separated.
Under this situation
we shall at first restrict ourselves to the case
$b=B=0$ since under the circumstances we can solve Eq. (5.4) exactly. 
Really we employ the ansatz (6.14) and obtain the system (due to the fact that 
$\sigma_1{\cal D}_0=-{\cal D}_0\sigma_1$) 
$$\left(\partial_r+
\frac{\lambda}{r}\right)F_{1}=
i(\mu_0-c)F_{2},$$
$$\left(\partial_r
-\frac{\lambda}{r}\right)F_{2}=
-i(\mu_0+c)F_{1} \>\eqno(6.16)$$
with an eigenvalue $\lambda$ for the eigenspinor $\Phi$ of the above
operator ${\cal D}_0$,
$\lambda=\pm(l+1)\in{\Bbb Z}\backslash\{0\}\>, l=0,1,2...$
(see previous Subsection). 

Let us now employ the ansatz
$$F_{1}=\sqrt{\mu_0-(\omega-gA)}\,r^{\alpha}e^{-\beta r}[f_{1}(x)+
f_{2}(x)],$$
$$F_{2}=i\sqrt{\mu_0+(\omega-gA)}\,r^{\alpha}e^{-\beta r}[f_{1}(x)-
f_{2}(x)]
\eqno(6.17)$$ 
with $\alpha=\sqrt{\lambda^2-g^2a^2}$, $\beta=
\sqrt{\mu_0^2-(\omega-gA)^2}$, $x=2\beta r$.

Then, inserting the ansatz into (6.16), adding and subtracting equations give
rise to
$$\beta xf_{1}'+Yf_{1}+Zf_{2}=0\>,\eqno(6.18a)$$
$$\beta xf_{2}'-\beta xf_{2}+Y_0f_{2}+Z_0f_{1}=0\>,\eqno(6.18b)$$
where prime signifies the differentiation with respect to $x$,
$Y,Y_0=\alpha\beta\mp ga(\omega-gA)$,
$Z,Z_0=\lambda\beta\pm ga\mu_0$. From (6.18), if using the relations 
$YY_0-ZZ_0=0$, $Y+Y_0=2\alpha\beta$, one yields the second order
equations in $x$
$$xf_{1}''+(1+2\alpha-x)f_{1}'-\frac{Y}{\beta}f_{1}=0\>,\eqno(6.19a)$$
$$xf_{2}''+(1+2\alpha-x)f_{2}'-
\left(1+\frac{Y}{\beta}\right)f_{2}=0\>,\eqno(6.19b)$$
that are the Kummer equations (confluent hypergeometric equations in another 
terminology, see, e. g. Ref.\cite{Abr64}) and for (6.19a) 
the only finite solution at 0 and at infinity not strongly increasing is the 
Laguerre polynomial $L^\rho_{n}(x)$ with $n=-Y/\beta=0, 1, 2,...$. 
This gives the spectrum
$$\omega=
gA\pm\mu_0\left[1+\frac{g^2a^2}{(n+
\sqrt{\lambda^2-g^2a^2})^2}\right]^{-1/2}\>,
\eqno(6.20)$$
wherefrom it is clear that constant $A$ only shift the origin of count for 
energy and we can consider $A=0$. 
Further, putting $f_1=CL^{2\alpha}_{n}(x)$ with some constant $C$, from (6.18a) 
at $n>0$ we find 
$$f_2=\frac{C}{Z}\left[\beta xL^{2\alpha+1}_{n-1}(x)-YL^{2\alpha}_{n}(x)
\right]$$ because 
$[L^{2\alpha}_{n}(x)]'=-L^{2\alpha+1}_{n-1}(x)$ \cite{Su79}, that 
entails
$$F_1=C\sqrt{\mu_0-\omega}\,r^{\alpha}e^{-\beta r}
\left[\left(1-\frac{Y}{Z}\right)L^{2\alpha}_{n}(x)+
\frac{\beta}{Z}xL^{2\alpha+1}_{n-1}(x)\right]\>,$$
$$F_2=iC\sqrt{\mu_0+\omega}\,r^{\alpha}e^{-\beta r}
\left[\left(1+\frac{Y}{Z}\right)L^{2\alpha}_{n}(x)-
\frac{\beta}{Z}xL^{2\alpha+1}_{n-1}(x)\right]\>.\eqno(6.21)$$
The case $n=0$ should be considered separately. We here have $Y=0$, $Y_0=
2\alpha\beta=2ga\omega$, 
$\omega=\pm\mu_0\sqrt{\lambda^2-g^2a^2}/|\lambda|$, $f_1=
CL^{2\alpha}_{0}(x)=C$. Further $ZZ_0=0=(\lambda\beta)^2-(ga\mu_0)^2$ which 
entails $|\lambda|\beta=g|a|\mu_0$.
Then at $a>0, \lambda>0$ we get $Z=\lambda\beta+ga\mu_0=
|\lambda|\beta+g|a|\mu_0=2ga\mu_0>0$, 
$Z_0=\lambda\beta-ga\mu_0=|\lambda|\beta-g|a|\mu_0=0$, $f_2=-CY/Z=
-CZ_0/Y_0=0$ and
$$F_1=C\sqrt{\mu_0-\omega}\,r^{\alpha}e^{-\beta r}\>,
F_2=iC\sqrt{\mu_0+\omega}\,r^{\alpha}e^{-\beta r}\>.\eqno(6.22)$$
At $a>0, \lambda<0$ we obtain 
$Z=\lambda\beta+ga\mu_0=-|\lambda|\beta+g|a|\mu_0=0>0$, $Z_0=-2ga\mu_0<0$,
$f_2=-CZ_0/Y_0=C\mu_0/\omega=C|\lambda|/(\pm\sqrt{\lambda^2-g^2a^2})$ and 
$$F_1=C\sqrt{\mu_0-\omega}\,r^{\alpha}e^{-\beta r}
\left(1+\frac{\mu_0}{\omega}\right)\>,
F_2=iC\sqrt{\mu_0+\omega}\,r^{\alpha}e^{-\beta r}
\left(1-\frac{\mu_0}{\omega}\right)\>.\eqno(6.23)$$
At $a<0, \lambda<0$ we get $Z=\lambda\beta+ga\mu_0=-|\lambda|\beta-g|a|\mu_0=
-2g|a|\mu_0<0$, $Z_0=\lambda\beta-ga\mu_0=-|\lambda|\beta+g|a|\mu_0 =0$, 
$f_2=0$ and $F_{1,2}$ are given by (6.22). At last, at $a<0, \lambda>0$ we 
have $Z=\lambda\beta+ga\mu_0=|\lambda|\beta-g|a|\mu_0=0$,
$Z_0=\lambda\beta-ga\mu_0=|\lambda|\beta+g|a|\mu_0 =2g|a|\mu_0$, 
$f_2=-CZ_0/Y_0=C\mu_0/\omega$ and $F_{1,2}$ are given by (6.23). 

To describe relativistic bound states we should require
$\Psi\in L_2^{4}({\Bbb R}^3)$ at any $t\in{\Bbb R}$ and one can
accept the normalization condition for
$F_{1}, F_{2}$ in the form
$$\int_0^\infty(|F_{1}|^2+|F_{2}|^2)dr=1\>\eqno(6.24)$$
with taking into account the condition (6.7) so that $C$ can be
determined from the relation (6.24).

It is the expression (6.20) that is in essence the Sommerfeld formula 
mentioned in Subsection 5.1. It should be noted, however, that standard 
parametrization in the Sommerfeld formula adduced in all the monographs 
(see, e. g., \cite{{LL1},{ST},{BD}}) is somewhat different from that in 
formula (6.20). The fact is that the standard approach uses
the orthonormal basis in $L_2^2({\Bbb S}^2)$ different from the basis 
of the eigenspinors of the Dirac operator ${\cal D}_0$. Namely, one uses 
the eigenbasis of the operator $K=\vec\sigma\vec L+1$, where 
$\vec\sigma=(\sigma_1,\sigma_2,\sigma_3)$ and 
$\vec L=-i(\vec r\times \partial/\partial\vec r)$ is the angular momentum 
operator. The eigenvalues of operator $K$ are numbers of the form 
$\lambda'=\pm(j+1/2)$ with $j=1/2,3/2,...$, so $j$ defines the total angular 
momentum $\vec J=\vec L+\vec S$ with spin momentum $\vec S=\vec\sigma/2$, 
that is $J^2$ has the eigenvalues $j(j+1)$. In standard approach 
just $\lambda'$ stands in (6.20) instead of $\lambda$. It is evident that 
both formulas (with $\lambda$ or $\lambda'$) reproduce the same spectrum but 
the corresponding wave functions will be slightly different in angular part 
depending on $\vartheta, \varphi$ since the Dirac 
operator ${\cal D}_0$ does not commute with operator $K$. There are at least 
two reasons why historically operator $K$ was employed rather than 
${\cal D}_0$. The first one is that in standard approach one solves Dirac 
equation (5.4) in Cartesian coordinates without transition to spherical ones and
under this situation there naturally arises just operator $K$ whereas when 
passing on to the spherical coordinates there naturally would arise just 
operator ${\cal D}_0$ while separating variables. The second reason is that 
the whole formula (6.20) is not necessary in nonrelativistic systems such as 
hydrogen atom, positronium and so on. It is enough to restrict themselves to 
a few terms of expansion in $g^2$ for (6.20) to obtain corrections (fine 
structure) to the nonrelativistic spectrum which are in concordance with 
experiment and one may use the notion of spin as essentially nonrelativistic 
phenomenon. Another matter are quarks in mesons that should probably  
be considered essentially relativistic objects and it is 
the most natural to write down Dirac equation (5.4) in Minkowski spacetime with 
spherical coordinates in its spatial part. It is the latter way that we pursue 
in present paper.

At the end of this Subsection we can slightly generalize the results obtained 
if inserting solution (5.19) at $b=B=0$ into Dirac equation (5.4). Then it is 
not complicated to see we shall get
the similar relations providing that $K=k/g$ with $k\in{\Bbb Z}$,
i. e., $k$ is integer number. But now we should consider the spinor
$\Phi$ of (6.14) to be
the eigenspinor of the twisted euclidean Dirac
operator ${\cal D}_k$ on the unit sphere ${\Bbb S}^2$ with
the Chern number $k$ (see Subsection 6.1) and the eigenvalues $\lambda$ 
should be, accordingly,
replaced by $\lambda=\pm\sqrt{(l+1)^2-k^2}$, $l\ge|k|$.
Physically the corresponding configurations
describe the Dirac monopole ones with magnetic charge
$P= k/g$ so the corresponding wave functions should be modified in 
obvious way. 

\subsection{$N=2$}
After inserting the solution (5.21) (with $A=0$) into (5.8) 
for both the equations we employ the ansatz
$$\Psi_j=e^{i\omega_j t}r^{-1}\pmatrix{F_{j1}(r)\Phi_j(\vartheta,\varphi)\cr\
F_{j2}(r)\sigma_1\Phi_j(\vartheta,\varphi)}\>,j=1,2 \eqno(6.25)$$
with a 2D spinor $\Phi_j=\pmatrix{\Phi_{j1}\cr\Phi_{j2}}$ which entails 
the systems
$$\left(\partial_r+
\frac{\lambda_j}{r}\right)F_{j1}=
i(\mu_0-c_j)F_{j2},$$
$$\left(\partial_r
-\frac{\lambda_j}{r}\right)F_{j2}=
-i(\mu_0+c_j)F_{j1} \>\eqno(6.26)$$
with an eigenvalue $\lambda_j$ for the eigenspinor $\Phi_j$ of the Dirac 
operator ${\cal D}_0$,
$\lambda_j=\pm(l_j+1)\in{\Bbb Z}\backslash\{0\}\>, l_j=0,1,2...$
Besides
$$c_1=\omega_1+ga/r,
c_2=\omega_2-ga/r\>,\eqno(6.27)$$
so that the energy spectrum $\omega$ of particle is given by the
relation $\omega=\omega_1+\omega_2$.
Acting along the same lines as in previous Subsection we obtain the spectrum 
of particle in the form
$$\frac{\omega}{\mu_0} =
\pm\left[1+\frac{g^2a^2}{(n_{1}+
\sqrt{\lambda_1^2-g^2a^2})^2}\right]^{-1/2}
\pm\left[1+\frac{g^2a^2}{(n_{2}+
\sqrt{\lambda_2^2-g^2a^2})^2}\right]^{-1/2}\>,\eqno(6.28)$$
where the number $n_{1,2}=0,1,2,...$.

If $K\ne0$ in (5.25) at $b=B=0$ then when inserting (5.21), (5.25) into (5.8) 
we shall get the similar spectrum (6.28) providing that $K=k/g$ with 
$k\in{\Bbb Z}$,
i. e., $k$ is an integer number. But now we should consider the spinor
$\Phi_j$ of (6.25) to be
the eigenspinor $\Phi_j$ of the twisted euclidean Dirac
operators ${\cal D}_{\pm k}$ on the unit sphere ${\Bbb S}^2$ 
(see Subsection 6.1), 
respectively, with the Chern numbers
$\pm k$ and the eigenvalues $\lambda_j$ should be, accordingly,
replaced by $\lambda_1=\pm\sqrt{(l_1+1)^2-k^2}$, $l_1\ge|k|$,
$\lambda_2=\pm\sqrt{(l_2+1)^2-k^2}$, $l_2\ge|k|$. 
Physically the corresponding configurations
of SU(2)-field describe the Dirac-like monopole ones with magnetic charges,
conformably, $P_1= k/g$, $P_2=-k/g$ but the total
(nonabelian) magnetic charge of the given configurations remains equal to
$P_1+P_2=0$.
 
The corresponding wave functions are not complicated to be written out on the 
analogy of U(1)-case of previous Subsection but we shall not dwell upon it. 
It should be only noted that the condtion (6.24) should be replaced by 
$$\int_0^\infty(|F_{j1}|^2+|F_{j2}|^2)dr=\frac{1}{2}\>,\, j=1,2.$$

\subsection{$N=3$}
We here use the solution for $A_t$ of (2.13) (with $A_1=A_2=0$) and for all 
three equations in (5.9) 
we employ the ansatz
$$\Psi_j=e^{i\omega_j t}r^{-1}\pmatrix{F_{j1}(r)\Phi_j(\vartheta,\varphi)\cr\
F_{j2}(r)\sigma_1\Phi_j(\vartheta,\varphi)}\>,j=1,2,3 \eqno(6.29)$$
with a 2D spinor $\Phi_j=\pmatrix{\Phi_{j1}\cr\Phi_{j2}}$ which entails 
the systems
$$\left(\partial_r+
\frac{\lambda_j}{r}\right)F_{j1}=
i(\mu_0-c_j)F_{j2},$$
$$\left(\partial_r
-\frac{\lambda_j}{r}\right)F_{j2}=
-i(\mu_0+c_j)F_{j1} \>\eqno(6.30)$$
with an eigenvalue $\lambda_j$ for the eigenspinor $\Phi_j$ of the Dirac 
operator ${\cal D}_0$,
$\lambda_j=\pm(l_j+1)\in{\Bbb Z}\backslash\{0\}\>, l_j=0,1,2...$
Besides
$$c_1=\omega_1+ga_1/r, c_2=\omega_2+ga_2/r, c_3=\omega_3-g(a_1+a_2)/r\>,
\eqno(6.31)$$
so that the energy spectrum $\omega$ of particle is given by the
relation $\omega=\omega_1+\omega_2+\omega_3$ which we obtain 
in the form
$$\frac{\omega}{\mu_0} =
\pm\left[1+\frac{g^2a_1^2}{(n_{1}+
\sqrt{\lambda_1^2-g^2a_1^2})^2}\right]^{-1/2}
\pm\left[1+\frac{g^2a_2^2}{(n_{2}+
\sqrt{\lambda_2^2-g^2a_2^2})^2}\right]^{-1/2}$$
$$\pm\left[1+\frac{g^2(a_1+a_2)^2}{(n_{3}+
\sqrt{\lambda_3^2-g^2(a_1+a_2)^2})^2}\right]^{-1/2}\>,\eqno(6.32)$$
where the number $n_{1,2,3}=0,1,2,...$.

If $K_j\ne0$ in (5.26) at $b_j=B_j=0$ then when inserting (2.13), (5.26) 
into (5.9) 
we shall get the similar spectrum (6.32) providing that $K_j=k_j/g$ with 
$k_j\in{\Bbb Z}$,
i. e., $k_j$ are integers. But now we should consider the spinor
$\Phi_j$ of (6.29) to be
the eigenspinor $\Phi_j$ of the twisted euclidean Dirac
operator ${\cal D}_k$ on the unit sphere ${\Bbb S}^2$ (see Subsection 6.1), 
respectively, with the Chern numbers
$k=k_1,k_2,-(k_1+k_2)$ and the eigenvalues $\lambda_j$ should be, accordingly,
replaced by $\lambda_1=\pm\sqrt{(l_1+1)^2-k_1^2}$, $l_1\ge|k_1|$,
$\lambda_2=\pm\sqrt{(l_2+1)^2-k_2^2}$, $l_2\ge|k_2|$,
$\lambda_3=\pm\sqrt{(l_3+1)^2-(k_1+k_2)^2}$, $l_3\ge|k_1+k_2|$.
The corresponding configurations
of gluonic field describe the Dirac-like monopole ones with magnetic charges,
conformably, $P_1= k_1/g$, $P_2=k_2/g$, $P_3=-(k_1+k_2)/g$, but the total
(nonabelian) magnetic charge of the given configurations remains equal to
$P_1+P_2+P_3=0$.
 
The corresponding wave functions are again not complicated to be written out 
on the analogy of U(1)-case of Subsection 6.2 but we do not dwell upon it and 
the condtion (6.24) should be replaced by 
$$\int_0^\infty(|F_{j1}|^2+|F_{j2}|^2)dr=\frac{1}{3}\>,j=1,2,3.$$

\subsection{$N=4$}
If inserting the solution for $A_t$ of (5.27) (with $A_1=A_2=A_3=0$) into 
(5.10) then for all four equations in (5.10) 
the ansatz of form (6.29) with $j=1,2,3,4$ will lead to the systems of form 
(6.30) with an eigenvalue $\lambda_j$ for the eigenspinor $\Phi_j$ of the Dirac 
operator ${\cal D}_0$, where 
$\lambda_j=\pm(l_j+1)\in{\Bbb Z}\backslash\{0\}\>, l_j=0,1,2...$, 
while
$$c_1=\omega_1+ga_1/r, c_2=\omega_2+ga_2/r, c_3=\omega_3+ga_3/r, 
c_4=\omega_4-g(a_1+a_2+a_3))/r\>,
\eqno(6.33)$$
so that the energy spectrum $\omega$ of particle is given by the
relation $\omega=\omega_1+\omega_2+\omega_3+\omega_4$ and is obtained  
in the form
$$\frac{\omega}{\mu_0} =
\pm\left[1+\frac{g^2a_1^2}{(n_{1}+
\sqrt{\lambda_1^2-g^2a_1^2})^2}\right]^{-1/2}
\pm\left[1+\frac{g^2a_2^2}{(n_{2}+
\sqrt{\lambda_2^2-g^2a_2^2})^2}\right]^{-1/2}$$
$$\pm\left[1+\frac{g^2a_3^2}{(n_{3}+
\sqrt{\lambda_3^2-g^2a_3^2})^2}\right]^{-1/2}
\pm\left[1+\frac{g^2(a_1+a_2+a_3)^2}{(n_{4}+
\sqrt{\lambda_4^2-g^2(a_1+a_2+a_3)^2})^2}\right]^{-1/2}\>,\eqno(6.34)$$
where the number $n_{1,2,3,4}=0,1,2,...$.

If $K_j\ne0$ in (5.28) at $b_j=B_j=0$ then when inserting (5.27), (5.28) 
into (5.10) 
we shall get the similar spectrum (6.34) providing that $K_j=k_j/g$ with 
$k_j\in{\Bbb Z}$,
i. e., $k_j$ are integers, so the spinor
$\Phi_j$ is 
the eigenspinor $\Phi_j$ of the twisted euclidean Dirac
operator ${\cal D}_k$ on the unit sphere ${\Bbb S}^2$ (see Subsection 6.1), 
respectively, with the Chern numbers
$k=k_1,k_2,k_3,-(k_1+k_2+k_3)$ and the eigenvalues $\lambda_j$ should be, 
accordingly, replaced by $\lambda_1=\pm\sqrt{(l_1+1)^2-k_1^2}$, $l_1\ge|k_1|$,
$\lambda_2=\pm\sqrt{(l_2+1)^2-k_2^2}$, $l_2\ge|k_2|$, 
$\lambda_3=\pm\sqrt{(l_3+1)^2-k_3^2}$, $l_3\ge|k_3|$
$\lambda_4=\pm\sqrt{(l_4+1)^2-(k_1+k_2+k_3)^2}$, $l_4\ge|k_1+k_2+k_3|$.
The corresponding configurations of SU(4)-field describe the Dirac-like 
monopole ones with magnetic charges,
conformably, $P_1= k_1/g$, $P_2=k_2/g$, $P_3=k_3/g$, 
$P_4=-(k_1+k_2+k_3)/g$ but as before the total
(nonabelian) magnetic charge of the given configurations remains equal to
$P_1+P_2+P_3+P_4=0$.
 
The corresponding wave functions are again not complicated to be written out 
on the analogy of U(1)-case of Subsection 6.2 but we do not dwell upon it 
while the condtion (6.24) should be replaced by 
$$\int_0^\infty(|F_{j1}|^2+|F_{j2}|^2)dr=\frac{1}{4}\>, j=1,2,3,4.$$

%\newpage
\section{Spectrum of Bound States in the Coulomb-Linear Case}
\subsection{U(1)-case}
We now should return to the system (6.15) at $b\ne0, B\ne0, A=0$. One may hope 
to obtain the almost exact solution of (6.15) if considering 
$\sigma_2\Phi\approx\sin{\vartheta}\,\Phi$. As follows from the estimate in 
Appendix $D$ this condition is rather good fulfiled and when doing so we probably 
make an error retaining eigenvalues $\lambda$ of the euclidean Dirac operator 
${\cal D}_0$ on the unit sphere ${\Bbb S}^2$ instead of the eigenvalues
of a less symmetric operator on ${\Bbb S}^2$ whose form is unknown explicitly. 

Having accepted the mentioned condition we come to the system (owing to the 
fact that $\sigma_1{\cal D}_0=-{\cal D}_0\sigma_1$, 
$\sigma_2\sigma_1=-\sigma_1\sigma_2$) 
$$\left[\partial_r+
\frac{\lambda}{r}-g\left(b+\frac{B}{r}\right)\right]F_{1}=
i(\mu_0-c)F_{2},$$
$$\left[\partial_r
-\frac{\lambda}{r}+g\left(b+\frac{B}{r}\right)\right]F_{2}=
-i(\mu_0+c)F_{1} \>\eqno(7.1)$$
with $c=\omega+ga/r$.

Now we employ the ansatz
$$F_{1}=Pr^{\alpha}e^{-\beta r}[f_{1}(x)+f_{2}(x)], 
F_{2}=iQr^{\alpha}e^{-\beta r}[f_{1}(x)-f_{2}(x)]\> \eqno(7.1')$$
with $\alpha=\sqrt{(\lambda-gB)^2-g^2a^2}$, $\beta=
\sqrt{\mu_0^2-\omega^2+g^2b^2}$, $P=gb+\beta$,
$Q=\mu_0+\omega$, $x=2\beta r$.

After this, inserting the ansatz into (7.1), adding and subtracting equations
entail
$$xPQf_{1}'+Yf_{1}+Zf_{2}=0\>,\eqno(7.2a)$$
$$xPQf_{2}'-xPQf_{2}+Y_0f_{2}+\left(Z_0-\frac{gb}{\beta}
PQx\right)f_{1}=0\>,
\eqno(7.2b)$$
where prime signifies the differentiation with respect to $x$,
$Y,Y_{0}=[\alpha\beta\mp ga\omega+g\alpha b]Q\pm g^2abP$,
$Z,Z_{0}=[(\lambda-gB)P\pm ga\mu_0)]Q\pm g^2abP$ and
$YY_0-ZZ_0=0$.

From (7.2$a$)--(7.2$b$) one yields the second order
equations in $x$
$$xf_{1}''+(1+2\alpha-x)f_{1}'
+nf_{1}=0\>,\eqno(7.3)$$
$$xf_{2}''+\left(\frac{Z_0}{Z_0-b_0x}+2\alpha-x\right)f_{2}'+
n\left(\frac{Z_0\kappa}{Z_0-b_0x}+1\right)f_{2}=0\>\eqno(7.4)$$
with $b_0=gbPQ/\beta$, $\kappa=PQ/Y$ and
$$n=\frac{gbZ-\beta Y}{\beta PQ}\>,\eqno(7.5)$$
which entails the equation for spectrum of $\omega$ 
$$[g^2a^2+(n+\alpha)^2]\omega^2+
2(\lambda-gB)g^2ab\,\omega+$$
$$[(\lambda-gB)^2-(n+\alpha)^2]g^2b^2-
\mu_0^2(n+\alpha)^2=0\>,  \eqno(7.6)$$
that yields
$$\omega=\frac{-(\lambda-gB)g^2ab\pm
\sqrt{X}}{g^2a^2+(n+\alpha)^2}  \eqno(7.7) $$
with $X=(\lambda-gB)^2g^4a^2b^2-[g^2a^2+(n+\alpha)^2]
\{[(\lambda-gB)^2-(n+\alpha)^2]g^2b^2-
\mu_0^2(n+\alpha)^2\}$ and
it is clear that the expression (7.7) 
passes on to (6.20) at $b=B=0$ while (7.7) can be rewritten in a more 
symmetrical form
$$\omega=\omega(n,l,\lambda)=
\frac{-\Lambda g^2ab\pm(n+\alpha)
\sqrt{(n^2+2n\alpha+\Lambda^2)\mu_0^2+g^2b^2(n^2+2n\alpha)}}
{n^2+2n\alpha+\Lambda^2}\>\eqno(7.8)$$
with $\Lambda=\lambda-gB=\pm(l+1)-gB$.
 
It is clear that according to (7.3) (which is the confluent hypergeometric 
equation) we should choose 
$f_{1}=CL^{2\alpha}_{n}(x)$ 
with the Laguerre polynomial $L^{2\alpha}_{n}(x)$ and some constant $C$ 
if $n=0,1,2...$. Then at $n>0$ from (7.2a) we find 
$$f_2=\frac{C}{Z}\left[xPQL^{2\alpha+1}_{n-1}(x)-YL^{2\alpha}_{n}(x)
\right]$$ because 
$[L^{2\alpha}_{n}(x)]'=-L^{2\alpha+1}_{n-1}(x)$ \cite{Su79}, that 
entails
$$F_1=CP\,r^{\alpha}e^{-\beta r}
\left[\left(1-\frac{Y}{Z}\right)L^{2\alpha}_{n}(x)+
\frac{PQ}{Z}xL^{2\alpha+1}_{n-1}(x)\right]\>,$$
$$F_2=iCQ\,r^{\alpha}e^{-\beta r}
\left[\left(1+\frac{Y}{Z}\right)L^{2\alpha}_{n}(x)-
\frac{PQ}{Z}xL^{2\alpha+1}_{n-1}(x)\right]\>.\eqno(7.9)$$
At $n=0$ we have $f_{1}=CL^{2\alpha}_{0}(x)=C$, $f_2=-CY/Z$, wherefrom 
$$F_1=CP\,r^{\alpha}e^{-\beta r}
\left(1-\frac{Y}{Z}\right)=
CP\,r^{\alpha}e^{-\beta r}
\left(1-\frac{gb}{\beta}\right)\>,$$
$$F_2=iCQ\,r^{\alpha}e^{-\beta r}
\left(1+\frac{Y}{Z}\right)=
iCQ\,r^{\alpha}e^{-\beta r}
\left(1+\frac{gb}{\beta}\right)
\>,\eqno(7.10)$$
inasmuch as $gbZ=\beta Y$ at $n=0$ according to (7.5).

Hence we can see that $\Psi$ of (6.14) $\in L_2^{4}({\Bbb R}^3)$ at any 
$t\in{\Bbb R}$ and, conformably, 
$\Psi$ may describe relativistic bound states 
with the energy spectrum (7.8). 

Also it should be noted that the influence of the Dirac monopole 
configurations for U(1)-field when $K\ne0$ in (5.19) can be treated by the 
same manner as in Subsection 6.2 if taking 
$\sigma_2\Phi\approx\sin{\vartheta}\Phi$ for
the eigenspinor $\Phi$ of the twisted euclidean Dirac
operator ${\cal D}_k$ on the unit sphere ${\Bbb S}^2$ with the conforming
Chern number $k$.

At last, constant $C$ from (7.9)--(7.10) is defined by the condition (6.24) 
as before.

\subsection{$N=2$}
The corresponding modifications of U(1)-case here are obvious so we shall only 
briefly describe them. The spectrum is given by $\omega=\omega_1+\omega_2$ with 
$$\omega_1=\omega_1(n_1,l_1,\lambda_1)=$$
$$\frac{-\Lambda_1 g^2ab\pm(n_1+\alpha_1)
\sqrt{(n_1^2+2n_1\alpha_1+\Lambda_1^2)\mu_0^2+g^2b^2(n_1^2+2n_1\alpha_1)}}
{n_1^2+2n_1\alpha_1+\Lambda_1^2}\>,\eqno(7.11)$$
$$\omega_2=\omega_2(n_2,l_2,\lambda_2)=$$
$$\frac{-\Lambda_2 g^2ab\pm(n_2+\alpha_2)
\sqrt{(n_2^2+2n_2\alpha_2+\Lambda_2^2)\mu_0^2+g^2b^2(n_2^2+2n_2\alpha_2)}}
{n_2^2+2n_2\alpha_2+\Lambda_2^2}\>\eqno(7.12)$$
with $\Lambda_1=\lambda_1-gB=\pm(l_1+1)-gB$, 
$\Lambda_2=\lambda_2+gB=\pm(l_2+1)+gB$,
$\alpha_1=\sqrt{\Lambda_1^2-g^2a^2}$, 
$\alpha_2=\sqrt{\Lambda_2^2-g^2a^2}$. The corresponding radial parts of 
wave functions (6.25) are given at $n_j=0$ by
$$F_{j1}=C_jP_jr^{\alpha_j}e^{-\beta_jr}\left(1-
\frac{Y_j}{Z_j}\right),F_{j2}=iC_jQ_jr^{\alpha_j}e^{-\beta_jr}\left(1+
\frac{Y_j}{Z_j}\right),\eqno(7.13)$$
while at $n_j>0$ by
$$F_{j1}=C_jP_jr^{\alpha_j}e^{-\beta_jr}\left[\left(1-
\frac{Y_j}{Z_j}\right)L^{2\alpha_j}_{n_j}(r_j)+
\frac{P_jQ_j}{Z_j}r_jL^{2\alpha_j+1}_{n_j-1}(r_j)\right],$$
$$F_{j2}=iC_jQ_jr^{\alpha_j}e^{-\beta_jr}\left[\left(1+
\frac{Y_j}{Z_j}\right)L^{2\alpha_j}_{n_j}(r_j)-
\frac{P_jQ_j}{Z_j}r_jL^{2\alpha_j+1}_{n_j-1}(r_j)\right]\eqno(7.14)$$
with the Laguerre polynomials $L^\rho_{n_j}(r_j)$, $r_j=2\beta_jr$,
where $\beta_j=\sqrt{\mu_0^2-\omega_j^2+g^2b^2}$,
$P_1=gb+\beta_1$, $P_2=-gb+\beta_2$, $Q_j=\mu_0+\omega_j$,
$Y_1=(\alpha_1\beta_1- ga\omega_1+g\alpha_1b)Q_1+ g^2abP_1$, 
$Y_2=(\alpha_2\beta_2+ ga\omega_2-g\alpha_2b)Q_2+ g^2abP_2$,
$Z_1=[(\lambda_1-gB)P_1+ga\mu_0]Q_1+ g^2abP_1$, 
$Z_2=[(\lambda_2+gB)P_2-ga\mu_0]Q_2+ g^2abP_2$. 
Also it should be noted 
that the quantum numbers $n_j$ are defined by the relations 
$$n_1=\frac{gbZ_1-\beta_1Y_1}{\beta_1P_1Q_1}, 
n_2=-\frac{gbZ_2+\beta_2Y_2}{\beta_2P_2Q_2},
\eqno(7.15)$$
Further, $C_j$ is determined
from the normalization condition
$$\int_0^\infty(|F_{j1}|^2+|F_{j2}|^2)dr=\frac{1}{2}\>.\eqno(7.16)$$
Consequently, we shall gain that in (6.25) 
$\Psi_j\in L_2^{4}({\Bbb R}^3)$ at any $t\in{\Bbb R}$ and, as a result,
$\Psi=(\Psi_1,\Psi_2)$ may describe relativistic bound states 
with the energy spectrum (7.11)--(7.12).

Finally, it should be noted that the influence of the Dirac-like monopole 
configurations for SU(2)-field when $K\ne0$ in (5.25) can be treated by the 
same manner as in Subsection 6.3 if taking 
$\sigma_2\Phi_j\approx\sin{\vartheta}\Phi_j$ for
the eigenspinor $\Phi_j$ of the twisted euclidean Dirac
operator ${\cal D}_{\pm k}$ on the unit sphere ${\Bbb S}^2$ with the 
conforming Chern numbers $\pm k$.

\subsection{$N=3$}
Let us adduce the coresponding results without going into details that can be 
easily reconstructed along the lines of Subsection 7.1. The spectrum is given 
by $\omega=\omega_1+\omega_2+\omega_3$ with 
$$\omega_1=\omega_1(n_1,l_1,\lambda_1)=$$ 
$$\frac{-\Lambda_1 g^2a_1b_1\pm(n_1+\alpha_1)
\sqrt{(n_1^2+2n_1\alpha_1+\Lambda_1^2)\mu_0^2+g^2b_1^2(n_1^2+2n_1\alpha_1)}}
{n_1^2+2n_1\alpha_1+\Lambda_1^2}\>,\eqno(7.17)$$
$$\omega_2=\omega_2(n_2,l_2,\lambda_2)=$$
$$\frac{-\Lambda_2 g^2a_2b_2\pm(n_2+\alpha_2)
\sqrt{(n_2^2+2n_2\alpha_2+\Lambda_2^2)\mu_0^2+g^2b_2^2(n_2^2+2n_2\alpha_2)}}
{n_2^2+2n_2\alpha_2+\Lambda_2^2}\>,\eqno(7.18)$$
$$\omega_3=\omega_3(n_3,l_3,\lambda_3)=$$
$$\frac{-\Lambda_3 g^2a_3b_3\pm(n_3+\alpha_3)
\sqrt{(n_3^2+2n_3\alpha_3+\Lambda_3^2)\mu_0^2+g^2b_3^2
(n_3^2+2n_3\alpha_3)}}
{n_3^2+2n_3\alpha_3+\Lambda_3^2}\>,\eqno(7.19)$$
where $a_3=-(a_1+a_2)$, $b_3=-(b_1+b_2)$, 
$\Lambda_j=\lambda_j-gB_j\>, j=1, 2, 3\>$, $B_3=-(B_1+B_2)$, 
$n_j=0,1,2,...$, while $\lambda_j=\pm(l_j+1)$ are
the eigenvalues of euclidean Dirac operator ${\cal D}_0$ 
on unit sphere with $l_j=0,1,2,...$. Besides
$$\alpha_1=\sqrt{\Lambda_1^2-g^2a_1^2}\>, 
\alpha_2=\sqrt{\Lambda_2^2-g^2a_2^2}\>, 
\alpha_3=\sqrt{\Lambda_3^2-g^2(a_1+a_2)^2}\>.
\eqno(7.20)$$
Further, radial parts of (6.29) are given at $n_j=0$ by (7.13) while 
at $n_j>0$ by (7.14). We have 
$r_j=2\beta_jr$,
where $\beta_j=\sqrt{\mu_0^2-\omega_j^2+g^2b_j^2}$ at $j=1,2,3$ with 
$b_3=-(b_1+b_2)$, 
$P_j=gb_j+\beta_j$, $j=1,2$, $P_3=-g(b_1+b_2)+\beta_3$, $Q_j=\mu_0+\omega_j$,
$Y_j=(\alpha_j\beta_j- ga_j\omega_j+g\alpha_jb_j)Q_j+ g^2a_jb_jP_j$, $j=1,2$,  
$Y_3=[\alpha_3\beta_3+ g(a_1+a_2)\omega_3-g\alpha_3(b_1+b_2)]Q_3+ 
g^2(a_1+a_2)(b_1+b_2)P_3$,
$Z_j=[(\lambda_j-gB_j)P_j+ga_j\mu_0]Q_j+ g^2a_jb_jP_j$, $j=1,2$, 
$Z_3=[(\lambda_3+g(B_1+B_2))P_3-g(a_1+a_2)\mu_0]Q_3+ 
g^2(a_1+a_2)(b_1+b_2)P_3$, quantum numbers $n_j$ are defined by the relations 
$$n_1=\frac{gb_1Z_1-\beta_1Y_1}{\beta_1P_1Q_1}, 
n_2=\frac{gb_2Z_2-\beta_2Y_2}{\beta_2P_2Q_2},
n_3=-\frac{g(b_1+b_2)Z_3+\beta_3Y_3}{\beta_3P_3Q_3}.
\eqno(7.21)$$
Further, $C_j$ of (7.13)--(7.14) is determined
from the normalization condition
$$\int_0^\infty(|F_{j1}|^2+|F_{j2}|^2)dr=\frac{1}{3}\>.\eqno(7.21')$$
As a consequence, we shall gain that in (6.29) 
$\Psi_j\in L_2^{4}({\Bbb R}^3)$ at any $t\in{\Bbb R}$ and, accordingly,
$\Psi=(\Psi_1,\Psi_2,\Psi_3)$ may describe relativistic bound states 
with the energy spectrum (7.17)--(7.19).

Finally, it should be noted that the influence of the Dirac-like monopole 
configurations for gluonic SU(3)-field when $K_j\ne0$ in (5.26) can be treated 
by the same manner as in Subsection 6.4 if taking 
$\sigma_2\Phi_j\approx\sin{\vartheta}\Phi_j$ for
the eigenspinor $\Phi_j$ of the twisted euclidean Dirac
operator ${\cal D}_ k$ on the unit sphere ${\Bbb S}^2$ with the 
conforming Chern numbers $k=k_1, k_2, -(k_1+k_2)$.

\subsection{$N=4$}
In line with previous Subsection it is not difficult to write out the results 
for the given case. The spectrum is given 
by $\omega=\omega_1+\omega_2+\omega_3+\omega_4$ with 
$$\omega_j=\omega_j(n_j,l_j,\lambda_j)=$$ 
$$\frac{-\Lambda_j g^2a_jb_j\pm(n_j+\alpha_j)
\sqrt{(n_j^2+2n_j\alpha_j+\Lambda_j^2)\mu_0^2+g^2b_j^2(n_j^2+2n_j\alpha_j)}}
{n_j^2+2n_j\alpha_j+\Lambda_j^2}\>, j=1,2,3\>,\eqno(7.22)$$
$$\omega_4=\omega_4(n_4,l_4,\lambda_4)=$$
$$\frac{-\Lambda_4 g^2a_4b_4\pm(n_4+\alpha_4)
\sqrt{(n_4^2+2n_4\alpha_4+\Lambda_4^2)\mu_0^2+g^2b_4^2
(n_4^2+2n_4\alpha_4)}}
{n_4^2+2n_4\alpha_4+\Lambda_4^2}\>,\eqno(7.23)$$
where $a_4=-(a_1+a_2+a_3)$, $b_4=-(b_1+b_2+b_3)$, 
$\Lambda_j=\lambda_j-gB_j\>, j=1, 2, 3, 4\>, 
B_4=-(B_1+B_2+B_3)\>,$
$n_j=0,1,2,...$, while $\lambda_j=\pm(l_j+1)$ are
the eigenvalues of euclidean Dirac operator ${\cal D}_0$ 
on unit sphere with $l_j=0,1,2,...$. Besides
$$\alpha_j=\sqrt{\Lambda_j^2-g^2a_j^2}\>, j= 1,2,3\>,
\alpha_4=\sqrt{\Lambda_4^2-g^2(a_1+a_2+a_3)^2}\>.
\eqno(7.24)$$
Further, radial parts of (6.29) are given at $n_j=0$ by (7.13) while 
at $n_j>0$ by (7.14). We have 
$r_j=2\beta_jr$,
where $\beta_j=\sqrt{\mu_0^2-\omega_j^2+g^2b_j^2}$ at $j=1,2,3,4$ with 
$b_4=-(b_1+b_2+b_3)$, 
$P_j=gb_j+\beta_j$, $j=1,2,3$, $P_4=-g(b_1+b_2+b_3)+\beta_4$, 
$Q_j=\mu_0+\omega_j$,
$Y_j=(\alpha_j\beta_j- ga_j\omega_j+g\alpha_jb_j)Q_j+ g^2a_jb_jP_j$, $j=1,2,3$,  
$Y_4=[\alpha_4\beta_4+ g(a_1+a_2+a_3)\omega_4-g\alpha_4(b_1+b_2+b_3)]Q_4+ 
g^2(a_1+a_2+a_3)(b_1+b_2+a_3)P_4$,
$Z_j=[(\lambda_j-gB_j)P_j+ga_j\mu_0]Q_j+ g^2a_jb_jP_j$, $j=1,2,3$, 
$Z_4=[(\lambda_4+g(B_1+B_2+B_3))P_4-g(a_1+a_2+a_3)\mu_0]Q_4+ 
g^2(a_1+a_2+a_3)(b_1+b_2+b_3)P_4$, quantum numbers $n_j$ are defined by the 
relations 
$$n_j=\frac{gb_jZ_j-\beta_jY_j}{\beta_jP_jQ_j}, j=1,2,3,
n_4=-\frac{g(b_1+b_2+b_3)Z_4+\beta_4Y_4}{\beta_4P_4Q_4}.
\eqno(7.25)$$
Further, $C_j$ of (7.13)--(7.14) is determined
from the normalization condition
$$\int_0^\infty(|F_{j1}|^2+|F_{j2}|^2)dr=\frac{1}{4}\>.\eqno(7.26)$$
Thus, in (6.29) 
$\Psi_j\in L_2^{4}({\Bbb R}^3)$ at any $t\in{\Bbb R}$ and, consequently,
$\Psi=(\Psi_1,\Psi_2,\Psi_3,\Psi_4)$ may describe relativistic bound states 
with the energy spectrum (7.22)--(7.23).   

Finally, it should be noted that the influence of the Dirac-like monopole 
configurations for gluonic SU(4)-field when $K_j\ne0$ in (5.28) can be treated 
by the same manner as in Subsection 6.5 if taking 
$\sigma_2\Phi_j\approx\sin{\vartheta}\Phi_j$ for
the eigenspinor $\Phi_j$ of the twisted euclidean Dirac
operator ${\cal D}_ k$ on the unit sphere ${\Bbb S}^2$ with the 
conforming Chern numbers $k=k_1, k_2,k_3, -(k_1+k_2+k_3)$.

\subsection{Remark about the case $N>4$}

As is not complicated to see, one can consider general case $N>4$. To obtain 
the corresponding confining solutions for given $N$ one should take a concrete 
realization of the conforming SU($N$)-Lie algebra whose form depends on $N$, as 
we have seen above. Then the explicit form of the solutions under discussion 
will also depend on the mentioned realization but it is clear that they will be 
characterized by real constants $a_j, b_j, A_j, B_j$, $j=1,...,N$ with  
$a_N=-(a_1+a_2+\cdots+a_{N-1})$, $b_N=-(b_1+b_2+\cdots+b_{N-1})$, 
$B_N=-(B_1+B_2+\cdots+B_{N-1})$ while one may put $A_j=0$. 

The shape of spectrum for 
relativistic bound states will obviously have the same form as in 
(7.22) with $\Lambda_j=\lambda_j-gB_j$, $n_j=0,1,2,...$, but $j=1,...,N$. The 
corresponding modifications for parameters $\alpha_j,\beta_j, P_j, Q_j$ of the 
wave functions in the form (7.13)--(7.14) are also evident. The normalization 
condition (7.26) will contain $1/N$ in its right-hand side. So 
$\Psi=(\Psi_1,\Psi_2,\ldots,\Psi_N)$ may describe relativistic bound states 
with the the energy spectrum $\omega=\omega_1+\omega_2+\cdots+\omega_N$. 
At last, the influence of the Dirac-like monopole 
configurations for gluonic SU($N$)-field can be treated 
by the same manner as in previous Sections.   

\subsection{Nonrelativistic Limit}
It makes sense to obtain the nonrelativistic limit (i.e. when 
$c\to\infty$) for spectrum (7.8) in order to 
us to have a possibility of estimating the contribution of relativistic 
effects. Changing $a\to a/(\hbar c)$, $B\to B/(\hbar c)$ and expanding (7.8) 
in $z=1/c$, we get
$$\omega(n,l,\lambda)=$$
$$\pm\mu_0c^2\left[1\mp
\frac{g^2a^2}{2\hbar^2(n+|\lambda|)^2}z^2\right]
-\left[\frac{\lambda g^2ab}{\hbar(n+|\lambda|)^2}\,
\pm\mu_0\frac{g^3Ba^2f(n,\lambda)}{\hbar^3(n+|\lambda|)^{13}}\right]z\,+
O(z^2)\>,\eqno(7.27)$$
where 
$$f(n,\lambda)=10n^9\lambda+120n^7\lambda^3+252n^5\lambda^5+120n^3\lambda^7+
10n\lambda^9+$$
$$\frac{|\lambda|}{\lambda}\left(n^{10}+45n^8\lambda^2+210n^6\lambda^4+
210n^4\lambda^6+45n^2\lambda^8+\lambda^{10}\right)\>.
\eqno(7.28)$$

As is seen from (7.27), at $c\to\infty$ the contribution of linear magnetic 
field to the spectrum really vanishes and spectrum becomes in essence purely 
Coulomb one (modulo the rest energy) which corresponds to the lines discussed 
in Subsection 5.1. 

\section{Uniqueness of the Confining Solutions}
\subsection{Uniqueness}
 Let us consider the question of uniqueness of the confining solutions.
The latter ones were defined in Section 1 and in Subsection 5.1 as the 
spherically symmetric solutions of the Yang-Mills 
equations ($B$.3) (at $J=0$) containing only the components of the 
SU($N$)-field which are Coulomb-like or linear in $r$. Additionally 
we impose the Lorentz condition ($B$.4) on the sought solutions. As was remarked 
in Appendix $B$, the latter condition is necessary for 
quantizing the gauge fields consistently within the framework of perturbation 
theory (see, e. g. Ref. \cite{Ryd85}), so we should impose the given condition. 
Let us for definiteness consider the case of group SU(3). 
Under this situation we can take the general ansatz of form 
$$A=r^\mu\mit\Gamma dt+A_rdr+A_\vartheta d\vartheta+
r^\nu\mit\Delta d\varphi \>,\eqno(8.1)$$
where $A_\vartheta= r^\rho T$ and matrices $\mit\Gamma=\alpha^a\lambda_a$, 
$\mit\Delta=\beta^a\lambda_a$, $T=\gamma^a\lambda_a$ with arbitrary real 
constants $\alpha^a, \beta^a, \gamma^a$. It could seem that 
there is a more general ansatz in the form 
$A=r^{\mu_a}\alpha^a\lambda_a dt+A_rdr+
r^{\rho_a}\gamma^a\lambda_a d\vartheta+
r^{\nu_a}\beta^a\lambda_a d\varphi$ but somewhat more complicated considerations 
than the ones below (see Appendix $E$) show that all the same we should have 
$\mu_a=\mu, \nu_a=\nu$ for any $a$ so, within the current Subsection,  
we at once consider this 
condition to be fulfilled to avoid unnecessary complications and also we put 
$\rho_a=\rho$ at any $a$ for simplicity though it is not obligatory 
(see Appendix $E$). 

For the ansatz (8.1) the Lorentz condition ($B.4$) takes the form
$$\partial_r(r^2\sin{\vartheta}g^{rr}A_r)+
\partial_\vartheta(r^2\sin{\vartheta}g^{\vartheta\vartheta}A_\vartheta)=0$$
which can be rewritten as 
$$\partial_\vartheta(\sin{\vartheta}A_\vartheta)+\sin{\vartheta}
\partial_r(r^2A_r)=0\>, \eqno(8.2)$$
wherefrom it follows $\cot{\vartheta}r^\lambda T+\partial_r(r^2A_r)=0$ while 
the latter entails 
$$A_r=\frac{C}{r^2}-\frac{\cot{\vartheta}r^{\rho-1}T}{\rho+1}\>
\eqno(8.3)$$
with a constant matrix $C$ belonging to SU(3)-Lie algebra. 
Then we can see that it should put $C=T=0$ or else $A_r$ will not be spherically 
symmetric and the confining one where only the powers of $r$ equal to $\pm1$ 
are admissible. As a result we come to the conclusion that one should put 
$A_r=A_\vartheta=0$ in (8.1).
After this we have $F=dA+gA\wedge A=-\mu r^{\mu-1}\mit\Gamma dt\wedge dr+ 
\nu r^{\nu-1}\mit\Delta dr\wedge d\varphi+gr^{\mu+\nu}
[\mit\Gamma,\mit\Delta]dt\wedge d\varphi$ which entails (with the help of 
($A$.6))   
$$\ast F=\mu r^{\mu+1}\sin{\vartheta}\mit\Gamma d\vartheta\wedge d\varphi-
\frac{\nu r^{\nu-1}\mit\Delta}{\sin{\vartheta}} dt\wedge d\vartheta-
\frac{gr^{\mu+\nu}}{\sin{\vartheta}}
[\mit\Gamma,\mit\Delta]dr\wedge d\vartheta$$ 
and the Yang-Mills equations ($B$.3) (at $J=0$) turn into
$$\mu(\mu+1)r^{\mu+1}\sin^2{\vartheta}\mit\Gamma=
g^2r^{\mu+2\nu}[\mit\Delta,[\mit\Gamma,\mit\Delta]]\>,$$
$$\nu(\nu-1)r^{\nu-2}\mit\Delta=
g^2r^{2\mu+\nu}[\mit\Gamma,[\mit\Gamma,\mit\Delta]]\>.
\eqno(8.4)$$
It is now not complicated to enumerate possibilities for obtaining the confining 
solutions in accordance with (8.4), where we should put $\mu=-1$, $\nu=1$.
\begin{enumerate}
\item $\mit\Gamma$=0 or $\mit\Delta$=0. This situation does obviously not 
correspond to a confining solution
\item $\mit\Gamma=C\mit\Delta$ with some constant $C$. This case conforms 
to that all the parameters $\alpha^a$ describing electric colour Coulomb field 
and the ones $\beta^a$ for linear magnetic colour field are proportional -- 
the situation is not quite clear from physical point of view
\item Matrices $\mit\Gamma,\mit\Delta$ are not equal to zero simultaneously and 
both matrices belong to Cartan subalgebra of SU(3)-Lie algebra. The parameters 
$\alpha^a, \beta^a$ of electric and magnetic colour fields are not connected 
and arbitrary, i. e. 
they should be chosen from experimental data. The given situation is the most 
adequate to the physics in question and the corresponding confining solution 
is in essence the same which has been obtained in Section 2 in the form 
(2.13)--(2.14). 
\end{enumerate}
One can somewhat generalize the starting ansatz (8.1) taking it in the form 
$A=(r^\mu\mit\Gamma +A')dt+(r^\nu\mit\Delta +B')d\varphi$ 
with matrices $A'=A^a\lambda_a$, $B'=B^a\lambda_a$ and constants $A^a,B^a$. 
Then considerations along the same above lines draw the conclusion that 
the nontrivial confining solution is described by 
$\mit\Gamma,\mit\Delta, A', B'$ belonging to Cartan subalgebra which we 
could already see in the solutions (2.13)--(2.14).

Clearly, the obtained results may be extended over all SU($N$)-groups and even 
over all semisimple compact Lie groups since for them the corresponding Lie 
algebras possess just the only Cartan subalgebra. Also we can talk about the 
compact non-semisimple groups, for example, U($N$). In the latter case 
additionally to Cartan subalgebra we have centrum consisting from the matrices 
of the form $\alpha I_N$ with arbitrary constant $\alpha$. Really we have 
obtained the confining solutions for U(1)-group in Subsection 2.2. The most 
relevant physical cases are of course U(1)- and SU(3)-ones (QED and QCD), 
therefore we shall not consider further generalizations of the results 
obtained. It should also be noted that the nontrivial confining solutions 
obtained exist at any gauge coupling constant $g$, i. e. they are 
essentially {\em nonperturbative} ones. At last, as we have seen in Sections 5, 6,
there exist the nontrivial confining solutions containing the Dirac-like 
monopole parts. The latter are, however, not the spherically symmetric ones and 
we shall here not dwell upon uniqueness for such solutions.

\subsubsection{Remark Concerning the Wilson Confinement Criterion}
In our recent paper \cite{Gon05} it was shown that the confining solutions 
under consideration satisfy the so-called Wilson confinement criterion 
formulated as far back as in Ref. \cite{Wil} (see also Ref. \cite{Ban81}).  
During a long time up to now this criterion remains the 
dominant one when building one or another approach to the confinement problem. 
As far as is known to us, however, so far no explicit solutions of the 
SU($N$)-Yang-Mills equations obeying the given criterion have been found.  So, 
taking into account uniqueness of the confininig solutions under discussion, 
perhaps there exist no other solutions meeting the mentioned criterion.

\subsection{ Nonrelativistic Confining Potentials}
As has been mentioned in Subsection 4.1, in meson spectroscopy 
(see, e.g., Refs. \cite{{GM},{Rob}} and references therein) one often uses
the nonrelativistic confining 
potentials. Those confining potentials between quarks here
are usually modelled in the form $a/r+br$ with some constants $a$ and $b$.
It is clear, however, that from the QCD point of view the interaction between
quarks should be described by the whole SU(3)-field $A_\mu=A^a_\mu T_a$,
genuinely relativistic object, the nonrelativistic potential being only some
component of $A^a_t$ surviving in the nonrelativistic limit when the light 
velocity $c\to\infty$. Let us explore whether such potentials may be the 
solutions of the Maxwell or SU(3)-Yang-Mills equations. Though this can be 
easily derived from the results obtained in Subsection 8.1 let us consider 
the given situation directly in view of its physical importance.

\subsubsection{Maxwell Equations}
      In the case of Maxwell equations ($B$.5) (at $J=0$) 
the ansatz $A=
A_tdt=(a/r+br)dt$ yields $F=dA=(a/r^2-b)dt\wedge dr$, 
$*F=\sin\vartheta(br^2-a)d\vartheta\wedge d\varphi$ and the relation
$d*F=2br\sin\vartheta dr\wedge d\vartheta\wedge d\varphi=0$ entails
$b\equiv0$. 
      
\subsubsection{SU(3)-Yang-Mills Equations}
We use the ansatz 
$$A=A^a_t\lambda_adt=(A'/r+B'r)dt\>\eqno(8.5)$$ 
with some constant matrices $A'=\alpha^a\lambda_a, B'=\beta^a\lambda_a$. Then 
$A\wedge A=0$, $F=dA+gA\wedge A=dA=(A'/r^2-B')dt\wedge dr$, 
$\ast F=\sin{\vartheta}(B'r^2-A')d\vartheta\wedge d\varphi$, $d\ast F= 
2B'r\sin{\vartheta}dr\wedge d\vartheta\wedge d\varphi$, 
$\ast F\wedge A-A\wedge\ast F=
-2[A',B']rdt\wedge d\vartheta\wedge d\varphi$. 
Under the circumstances the Yang-Mills equations ($B.3$) (at $J=0$) are 
tantamount to the conditions 
$d\ast F=0$, $\ast F\wedge A-A\wedge\ast F=0$. The former entails 
$B'=0$, then the latter is fulfilled at any $A'$ and we can see that the 
Coulomb-like field $A=(A'/r)dt$ is a solution of the 
Yang-Mills equations ($B.3$) (at $J=0$) with arbitrary constant matrix $A'$  
which actually has been obtained in Subsection 8.1. 
In principle the ansatz (8.5) might be a solution of ($B.3$) with the source 
of the form
$$J=2B'r\sin{\vartheta}dr\wedge d\vartheta\wedge d\varphi+
2g[A',B']rdt\wedge d\vartheta\wedge d\varphi=$$
$$\ast j=\ast(j^a_\mu\lambda_a dx^\mu)=\ast\left(
\frac{2B'}{r}dt+g\frac{2[A',B']}{r\sin{\vartheta}}dr\right)\>,\eqno(8.6)$$
but ${\rm div}(j)\ne0$ and this is not consistent with the only source (5.6) 
derived from the QCD lagrangian (5.1). We can avoid this difficulty putting 
matrices $A', B'$ are not equal to zero simultaneously and both matrices 
belong to Cartan subalgebra of SU(3)-Lie algebra. Then 
$B'=\beta_3\lambda_3+\beta_8\lambda_8$ and for consistency with the only 
admissible source of (5.6) we should require source of (5.6) to be equal to 
one of (8.6) which entails 
$$g\overline{\Psi}(\gamma_\mu\otimes I_3)\lambda^a\Psi\lambda_a\,dx^\mu=
\frac{2(\beta^3\lambda_3+\beta^8\lambda_8)}{r}dt\>,$$ 
wherefrom one can conclude that 
$$g\overline{\Psi}(\gamma_t\otimes I_3)\lambda^a\Psi=0, a\ne3, 8\>,
g\overline{\Psi}(\gamma_t\otimes I_3)\lambda^3\Psi=\frac{2\beta^3}{r}\>,$$
$$ g\overline{\Psi}(\gamma_t\otimes I_3)\lambda^8\Psi=\frac{2\beta^8}{r}\>,
g\overline{\Psi}(\gamma_\mu\otimes I_3)\lambda^a\Psi=0, a=1,...,8,\,
\mu\ne t\>,\eqno(8.7)$$
which can obviously be satisfied only at 
$\beta^3\sim\beta^8\sim \Psi\to 0$ at each point of Minkowski spacetime, i. e., 
really matrix $B'=0$ again. All the above can easily be generalized to 
any $N>1$.

As a result, the potentials employed in nonrelativistic approaches do 
not obey the Maxwell or Yang-Mills equations.
The latter ones are essentially relativistic and, as we have seen, the
components linear in $r$ of the whole $A_\mu$ are different from $A_t$ and
related with magnetic (colour) field vanishing in the nonrelativistic limit.

\subsection{Remark about Search for Nonrelativistic Confining Potentials} 
The above results make us cast a new glance at search of many years for 
nonrelativistic potentials modelling the confinement. Many efforts were 
devoted to the latter topic, for example, within the framework of lattice 
gauge theories or potential approach (see, e.g., Refs. \cite{Pot} and references 
therein). It should be noted, however, that almost in all literature on this 
direction one does not bring up a question: whether such potentials could 
(or should) satisfy the Yang-Mills equations? As is clear from the above the 
answer is negative. That is why the mentioned approaches seem to be 
inconsistent. 

%\section{Diagonal gauge and the Frobenius theorem}
\section{Application to the Charmonium Spectrum}
As we have emphasized in Subsection 5.1, all considerations in the given paper 
develop with the aim of further physical applications or else the results 
obtained could be only of academic interest. Though there are no obstacles to 
apply the results (in the most physically interesting case $N=3$) to any 
meson, up to the moment of writing this paper all applications were concentrated 
on quarkonia (charmonium and bottomonium) and the results are contained in 
Refs. \cite{{Gon03},{GC03},{Gon04}}. Referring for more details 
to those papers, we shall here outline only the main conclusions drawn from the 
mentioned papers which confirm the physical picture underlying considerations 
of the given paper and warrant the linear confinement scenario described in 
Subsection 4.2.

\subsection{Relativistic Spectrum of Charmonium}
We can use (7.17)--(7.19) with various combinations of signes ($\pm$) before 
second summand in numerators of those formulas. In 
Refs. \cite{{Gon03},{GC03},{Gon04}} due to some reasons (inessential now) the 
combination ($++-$) was employed and besides the replacement 
$\omega_2\to\omega_3$ was made. Let us rewrite, e. g., the results 
out of Ref. \cite{Gon04} about spectrum according to (7.17)--(7.19) but keeping 
the mentioned combination ($++-$). Then numerical results for constants 
parametrizing the charmonium spectrum are shown in Table 1.

\begin{table}[htbp]
\caption{Gauge coupling constant, mass parameter $\mu_0$ and
parameters of the confining SU(3)-gluonic field for charmonium.}
\label{t.1}
\begin{center}
\begin{tabular}{|c|c|c|c|c|c|c|c|}
\hline
%\noalign{\hrule}\\
\small $ g$ & \small $\mu_0$ & \small $a_1$  & \small $a_2$ & \small $b_1$ 
& \small $b_2$ & \small $B_1$ & \small $B_2$ \\
  & \small (GeV) &  &  & \small (GeV) & \small (GeV) & &  \\
\hline
%\noalign{\hrule}\\
\small 0.46900 & \small 0.62500 & \small 2.21104 & \small -0.751317 
& \small 20.2395 & \small -12.6317 & \small 6.89659 & \small  6.89659 \\
%\noalign{\hrule}\\
\hline
\end{tabular}
\end{center}
\end{table}
One can note that the obtained mass parameter $\mu_0$ is consistent with the
present-day experimental limits \cite{pdg} where the current mass of $c$-quark
($2\mu_0$) is accepted between 1.1 GeV and 1.4 GeV. As for the gauge coupling 
constant $g$ then its value has been chosen in accordance with many recent 
considerations \cite{gc}, wherefrom one can conclude that the strong coupling
constant $\alpha_s=g^2$ is of order $0.22\approx0.469^2$ at the scale of the 
$c$-quark current mass. 
At last, as to parameters $A_{1,2}$ of solution (2.13), then they 
only shift the origin of count for the corresponding energies
and we can consider $A_1=A_2=0$, as was mentioned in Subsections 6.2, 6.4. 

With the constants of Table 1 the present-day levels of charmonium spectrum
were calculated with the help of (7.17)--(7.19) so Table 2 contains experimental 
values of these levels (from Ref. \cite{pdg})
and our theoretical ones computed according to the shown combinations
(we use the notations of levels from Ref. \cite{pdg}).
\begin{table}[htbp]
\centering
\caption{Theoretical and experimental charmonium levels.}
\label{t.2}
%\begin{center}
\begin{tabular}{|l|l|l|} 
\hline
\small State & 
\small Theoret. energy $\varepsilon_j=\sum\limits_{k=1}^3\omega_k$ &  
\small Experim. \\
 & (GeV) & value \\
 &  & (GeV) \\
\hline
\small $\eta_c(1S)$ & \small $\varepsilon_1= \omega_1(0,0,-1)+\omega_2(0,0,-1)+
\omega_3(0,0,-1)= 2.979597$ & \small 2.979600 \\
\hline
\small $J/\psi(1S)$ & \small $\varepsilon_2= \omega_1(0,0,-1)+\omega_2(0,0,-1)+
\omega_3(0,0,1)= 3.096913$ & \small 3.096916 \\ 
\hline
\small $\chi_{c0}(1P)$ & \small $\varepsilon_3= \omega_1(0,0,-1)+\omega_2(0,0,1)+
\omega_3(0,0,-1)= 3.415186$  & \small 3.415190 \\
\hline
\small $\chi_{c1}(1P)$ & \small
$\varepsilon_4= \omega_1(0,0,1)+\omega_2(0,1,-1)+\omega_3(2,0,1)
= 3.505304 $ & \small 3.510590 \\
\hline
\small $h_{c}(1P)$ & \small
$\varepsilon_5= \omega_1(0,0,-1)+\omega_2(0,0,1)+\omega_3(0,0,1)
= 3.532503$ & \small 3.526210 \\
\hline
\small $\chi_{c2}(1P)$ & \small
$\varepsilon_6= \omega_1(0,1,-1)+\omega_2(1,1,-1)+\omega_3(1,1,-1)
= 3.553097$ & \small 3.556260 \\
\hline
\small $\eta_c(2S)$ & \small
$\varepsilon_7= \omega_1(0,0,1)+\omega_2(0,1,-1)+\omega_3(1,0,-1)
= 3.671608$ & \small 3.65400 \\
\hline
\small $\psi(2S)$ & \small
$\varepsilon_8= \omega_1(0,1,-1)+\omega_2(1,1,-1)+\omega_3(2,1,1)
= 3.674025$ & \small 3.685093 \\
\hline
\small $\psi(3770)$ & \small
$\varepsilon_9= \omega_1(0,0,1)+\omega_2(0,0,1)+\omega_3(2,0,-1)
= 3.775598$ & \small 3.770000\\
\hline
\small $\psi(3836)$ & \small
$\varepsilon_{10}= \omega_1(0,1,-1)+\omega_2(0,1,1)+\omega_3(0,0,1)
= 3.833640$ & \small 3.836000\\
\hline
\small $X(3872)$ & \small
$\varepsilon_{11}= \omega_1(0,1,-1)+\omega_2(0,1,1)+\omega_3(0,1,1)
= 3.871672$ & \small 3.872000\\
\hline
\small $\psi(4040)$ & \small
$\varepsilon_{12}= \omega_1(0,0,1)+\omega_2(0,1,-1)+\omega_3(1,1,1)
= 4.042660$ & \small 4.040000\\
\hline
\small $\psi(4160)$ & \small
$\varepsilon_{13}= \omega_1(0,0,-1)+\omega_2(0,1,1)+\omega_3(0,0,-1)
= 4.153765$ & \small 4.159000 \\
\hline
\small $\psi(4415)$ & \small
$\varepsilon_{14}= \omega_1(0,0,-1)+\omega_2(1,0,-1)+\omega_3(2,1,1)
= 4.409260$ & \small 4.415000 \\
\hline
\end{tabular}
%\end{center}
\end{table}

\subsection{Nonrelativistic Limit}
One can be interested in estimating the contribution of relativistic effects, 
i. e. those connected with magnetic colour field linear in $r$. This has 
been done in  Refs. \cite{{Gon03},{GC03}} with the help of relations 
(7.27)--(7.28). The contribution of relativistic
effects can amount to tens per cent and they cannot be considered 
small. Moreover, the more excited the state of charmonium the worse the 
nonrelativistic approximation. The physical reason of it is quite clear. Really, 
we have seen in the nonrelativistic limit (see the relations [7.27)--(7.28)]  
that the parameters $b_{1,2}, B_{1,2}$ of the linear interaction between 
quarks vanish under this limit and the
nonrelativistic spectrum is independent of them and is practically getting
the pure Coulomb one. As a consequence, the picture of linear confinement for
quarks should be considered an essentially relativistic one while
the nonrelativistic limit is very crude approximation. In fact, as 
follows from exact solutions of SU(3)-Yang--Mills equations of (2.13), the 
linear interaction between quarks is connected with magnetic colour field that
dies out in the nonrelativistic limit, {\it i.e.} for static quarks. Only for
the moving rapidly enough quarks the above field will appear and generate
linear confinement between them. So the spectrum will depend on both
the static Coulomb electric colour field and the dynamical magnetic colour 
field responsible for the linear confinement for quarks which is just confirmed
by the relations (7.17)--(7.19). In our case, the interaction effect with the 
magnetic colour field is taken into consideration from the very outset which 
just reflects the linear confinement at large distances.

\subsection{Electromagnetic Transitions}
We can specify the obtained above charmonium spectrum. The fact is 
that the relations (7.17)--(7.19) permit various parametrizations of the 
charmonium spectrum (see Refs. \cite{{Gon03},{GC03}}) and therefore it should 
impose further conditions to fix a certain parametrization among several possible 
ones. For example, one can compute widths of elecromagnetic transitions among 
levels of charmonium, in particular for transitions 
$J/\psi(1S)\to \eta_c(1S)+\gamma$ and $\chi_{c0}(1P)\to J/\psi(1S)+\gamma$. In 
Ref. \cite{Gon04} the widths of the latter transitions have been calculated 
in dipole approximation that allowed one to use the corresponding wave 
functions described in Subsection 7.3. Results of computation supplies us with 
an additional justification for the choice of parameters 
of the SU(3)-confining gluon field adduced in Table 1 and allows us to conclude 
that dipole approximation is not enough for the second transition of the ones 
under discussion. The question now is what gluon concentrations are in the 
mentioned SU(3)-confining gluon field.

\subsection{Estimates of Gluon Concentrations and Magnetic Colour Field 
Sstrength}
To obtain necessary estimates we shall use an analogy with classical 
electrodynamics where is well known (see e. g. \cite{LL}) that the notion of 
classical electromagnetic field (a photon condensate) generated by a charged 
particle is applicable only at distances $>>$ the Compton wavelength 
$\lambda_c=1/m$ for the given particle. If denoting $\lambda_B$ the de Broglie 
wavelength of the particle then $\lambda_B=1/p$ with the relativistic impulse
$p=mv/\sqrt{1-v^2}$ while $v$ is the velocity of the particle (as a 
result, $\lambda_c=\lambda_B$ at $v=1/\sqrt{2}$) so one can 
rewrite $\lambda_B=\lambda_c\sqrt{1-v^2}/v$ and it is clear that 
$\lambda_B\to0$ when $v\to1$ (ultrarelativistic case), i. e., the particle 
becomes more and more point-like one. Accordingly, one can 
conclude that in the latter case the notion of classical electromagnetic field 
generated by a charged ultrarelativistic particle is applicable at distances 
$>>\lambda_B$. Under the circumstances, if the ultrarelativistic charged 
particle accomplishes its motion within the region with characteristic 
size of order $r_0$ then in the given region the electromagnetic field 
generated by the particle may be considered as classical one at
$r_0>>\lambda_B$. For example, in the case of positronium we have $r_0\sim
2a_0$, where $a_0\approx\ 5.29\cdot10^{-11}\ {\rm m}$ is the Bohr radius, so
$r_0>>\lambda_e\approx\ 3.86\cdot10^{-13}\ {\rm m}$, the electron Compton 
wavelength, i. e., the electric Coulomb interaction between electron and
positron in positronium can be considered classical electromagnetic field.
  
Passing on to QCD, gluons and quarkonia, it should be noted that quarks 
in quarkonia accomplish a finite motion within 
a region of order $1\ {\rm fm}=10^{-15}\ {\rm m}$. Then, as 
is seen from the radial parts of the wave functions described in Subsection 
7.3, the quantity 
$1/\beta_j$ permits to be considered
a characteristic size of the $j$-th colour component of the given quarkonium 
state and, consequently, we can take the magnitude 
$$r_0=\frac{1}{3}\sum_{j=1}^{3}\frac{1}{\beta_j}\>\eqno(9.1)$$
for a characteristic size of the whole quarkonium state and, in line with 
the above, we should consider the confining SU(3)-gluonic Yang-Mills field of
(2.13) or (2.14) to be classical one when $r_0>>\lambda_B$, the de Broglie 
wavelength of the corresponding quarks forming quarkonium.

  On the other hand, a classical electromagnetic field (photon condensate) 
conforms to the large photon concentrations for every frequency 
presented in the field \cite{LL1}. Then in QCD we should require the large 
gluon concentrations in the given classical gluonic field (gluon condensate). 
For the necessary estimates we shall employ the $T_{00}$-component of 
the energy-momentum tensor for a SU($N$)-Yang-Mills field
$$T_{\mu\nu}={1\over4\pi}\left(-F^a_{\mu\alpha}\,F^a_{\nu\beta}\,g^{\alpha\beta}+
{1\over4}F^a_{\beta\gamma}\,F^a_{\alpha\delta}g^{\alpha\beta}g^{\gamma\delta}
g_{\mu\nu}\right)\>. \eqno(9.1') $$

To estimate the given concentrations
we can employ $T_{00}$-component (volumetric energy density) of the 
energy-momentum tensor of $(9.1')$ and, taking the quantity 
$\omega= \mit\Gamma$, the whole decay width of the quarkonium state, for 
the characteristic frequency we obtain
the sought characteristic concentration $n$ in the form
$$n=\frac{T_{00}}{\mit\Gamma}\>. \eqno(9.2)$$

It is not complicated to obtain the curvature matrix (field strentgh) 
corresponding to the solution (2.13) or (2.14)
$$F= F^a_{\mu\nu}\lambda_a dx^\mu\wedge dx^\nu=
-\partial_r(A^a_t\lambda_a)dt\wedge dr
+\partial_r(A^a_\varphi \lambda_a)dr\wedge d\varphi
\>, \eqno(9.3)$$
which entails the only nonzero components
$$F^3_{tr}=\frac{a_1-a_2}{2r^2},\>
F^8_{tr}=\frac{(a_1+a_2)\sqrt{3}}{2r^2},\>
F^3_{r\varphi}=\frac{b_1-b_2}{2},\>
F^8_{r\varphi}=\frac{(b_1+b_2)\sqrt{3}}{2}\>\eqno(9.4) $$
and, in its turn,
$$T_{00}\equiv T_{tt}=\frac{1}{4\pi}\left\{\frac{3}{4}\left[(F^3_{tr})^2+
(F^8_{tr})^2\right]
+\frac{1}{4r^2\sin^2{\vartheta}}\left[(F^3_{r\varphi})^2+
(F^8_{r\varphi})^2\right]\right\}= $$
$$\frac{3}{16\pi}\left(\frac{a_1^2+a_1a_2+a_2^2}{r^4}+
\frac{b_1^2+b_1b_2+b_2^2}{3r^2\sin^2{\vartheta}}\right),\>\eqno(9.5)$$
so, further putting $\sin^2{\vartheta}=1/3$ for simplicity, we can rewrite
(9.5) in the form
$$T_{00}=T_{00}^{\rm coul}+T_{00}^{\rm lin}\>\eqno(9.6)$$
conforming to the contributions from the Coulomb and linear parts of the
solutions (2.13) or (2.14). The latter gives the corresponding split of $n$ from 
(9.2)
$$n=n_{\rm coul} + n_{\rm lin}.\>\eqno(9.7)$$
Using the Hodge star operator in 3-dimensional euclidean space [where 
$ds^2=g_{\mu\nu}dx^\mu\otimes dx^\nu\equiv 
dr^2+r^2(d\vartheta^2+\sin^2\vartheta d\varphi^2)$, 
$\ast(dr\wedge d\vartheta)=\sin{\vartheta}d\varphi, 
\ast(dr\wedge d\varphi)=-d\vartheta/\sin{\vartheta},
\ast(d\vartheta\wedge d\varphi)=dr/(r^2\sin{\vartheta}$], 
we can confront components $F^3_{r\varphi},F^8_{r\varphi}$ with 3-dimensional 
1-forms of the magnetic colour field 
$${\bf H}^3=-\frac{b_1-b_2}{2\sin{\vartheta}}d\vartheta\>,
{\bf H}^8=-\frac{(b_1+b_2)\sqrt{3}}{2\sin{\vartheta}}d\vartheta $$ 
that are modulo equal to 
$\sqrt{g^{\mu\nu}{\bf H}^{3,8}_{\mu}{\bf H}^{3,8}_{\nu}}$ or 
$$H^3=\frac{|b_1-b_2|}{2r\sin{\vartheta}},\>
H^8=\frac{|b_1+b_2|\sqrt{3}}{2r\sin{\vartheta}}\>,\eqno(9.8)$$
where we also consider $\sin^2{\vartheta}=1/3$ for simplicity.
 
For comparison we shall also estimate the photon concentration in the ground 
state of the positronium. As is known historically \cite{Per}, the analogy 
between the latter 
system and quarkonia played the important part when building the quarkonia 
models. For positronium we have the electromagnetic Coulomb interaction
$A=A_tdt=(e/r)dt$ which entails $F=F_{tr}dt\wedge dr=(e/r^2)dt\wedge dr$ and
$$T_{00}\equiv T_{tt}=\frac{1}{4\pi}(F_{tr})^2=\frac{\alpha_{em}}{4\pi r^4}
\>\eqno(9.9)$$
with $\alpha_{em}=e^2=1/137.0359895$.

\subsection{Numerical Results and Concluding Remarks}
When computing for the ground state of charmonium 
we used its present-day whole decay width 
${\mit\Gamma}= 17.3\ {\rm MeV}$ \cite{pdg}, while the calculation 
$r_0$ of (9.1) gives $r_0=r_1$ (see Table 3).

In the positronium case we employed the widths ${\mit\Gamma}_0=1/\tau_0$ 
(parapositronium) and ${\mit\Gamma}_1=1/\tau_1$ (orthopositronium), 
respectively, with the life times $\tau_0=1.252\cdot10^{-10}\ {\rm s}$, 
$\tau_1=1.377\cdot10^{-7}\ {\rm s}$ \cite{Per} while $r_0=2a_0$ with the
Bohr radius $a_0=0.529177249\cdot10^{5}\ {\rm fm}$ \cite{pdg}.

Tables 3, 4 contain the numerical results for both the cases, where,  
when calculating, we applied the relations  
$1\ {\rm GeV^{-1}}\approx0.21030893\ {\rm fm}\>$, 
$1\ {\rm T}\approx0.692508\times 10^{-15}\ {\rm GeV}^2$.

\begin{table}[htbp]
\caption{Gluon concentrations and magnetic colour field strengths in the 
ground state of charmonium.}
\label{t.3}
\begin{center}
\begin{tabular}{|llllll|}
%\hline
%\noalign{\hrule}\\
\hline
\scriptsize $\eta_c(1S)$: & \scriptsize $r_1= 0.0399766\ {\rm fm}$ &  & &  &\\
\hline 
%\noalign{\hrule}\\
\tiny $r$ & \tiny $n_{\rm coul}$ & \tiny $n_{\rm lin}$ 
& \tiny $n$ & \tiny $H^3$ & \tiny $H^8$ \\
\tiny (fm) & \tiny $ ({\rm m}^{-3}) $ 
& \tiny (${\rm m}^{-3}) $ & \tiny (${\rm m}^{-3}) $ 
& \tiny $({\rm T})$ & \tiny $({\rm T})$\\
\hline
%\noalign{\hrule}\\
\tiny $0.1r_1$ & \tiny $0.550649\cdot10^{58}$   
& \tiny $0.727630\cdot10^{55}$ & \tiny
$0.551377\cdot10^{58}$ & \tiny $0.216259\cdot10^{19}$  
& \tiny $0.866919\cdot10^{18}$ \\
\hline
\tiny$r_1$ & \tiny$0.550649\cdot10^{54}$ & \tiny$0.727630\cdot10^{53}$ 
& \tiny$0.623412\cdot10^{54}$ 
& \tiny$0.216259\cdot10^{18}$  & \tiny$0.866919\cdot10^{17}$  \\
\hline
\tiny$10r_1$ & \tiny$0.550649\cdot10^{50}$  & \tiny$0.727630\cdot10^{51}$ 
& \tiny$0.782695\cdot10^{51}$ 
& \tiny$0.216259\cdot10^{17}$  & \tiny$0.866919\cdot10^{16}$  \\
\hline
\tiny$1.0$ & \tiny$0.140637\cdot10^{49}$  & \tiny$0.116285\cdot10^{51}$ & 
\tiny$0.117691\cdot10^{51}$ & \tiny$0.864530\cdot10^{16}$  
& \tiny$0.346565\cdot10^{16}$  \\
\hline
\tiny$a_0$ & \tiny$0.179347\cdot10^{30}$  & \tiny$0.415260\cdot10^{41}$ & 
\tiny$0.415260\cdot10^{41}$ & \tiny$0.163372\cdot10^{12}$ 
& \tiny$0.654912\cdot10^{11}$  \\
\hline
%\noalign{\hrule}\\
\end{tabular}
\end{center}
\end{table}

\begin{table}[htbp]
\caption{Photon concentrations in the ground state of positronium.} 
\label{t.4}
\begin{center}
\begin{tabular}{|lll|}
\hline
%\noalign{\hrule}\\
   & \scriptsize $r_0= 2a_0=2\cdot0.529177249\cdot10^{5} 
\scriptsize$\ {\rm fm} &  \\
\hline
%\noalign{\hrule}\\
  & \scriptsize Parapositronium & \scriptsize Orthopositronium   \\
\scriptsize $r$ &  \scriptsize $n_{\rm para}$ & \scriptsize $n_{\rm ortho}$ \\
\scriptsize (fm) & \scriptsize(${\rm m}^{-3}$) &
 \scriptsize (${\rm m}^{-3}$)  \\
\hline
%\noalign{\hrule}\\
\tiny $0.01r_0$ & \tiny$0.888025\cdot10^{46}$  & 
\tiny$0.976685\cdot10^{49}$   \\
\hline
\tiny$0.1r_0$ & \tiny$0.888025\cdot10^{42}$   & 
\tiny$0.976685\cdot10^{45}$ \\
\hline
\tiny$r_0$ & \tiny$0.888025\cdot10^{38}$  & 
\tiny$0.976685\cdot10^{41}$    \\
\hline
\tiny$2r_0$ & \tiny$0.555015\cdot10^{37}$ & 
\tiny$0.610428\cdot10^{40}$   \\
\hline
%\noalign{\hrule}\\
\end{tabular}
\end{center}
\end{table}

 Then, as is seen from Tables 3, 4, qualitative 
behaviour of both the concentrations is similar. At the characteristic scales
of each system the concentrations are large and the corresponding fields 
(electric and magnetic colour ones or electric Coulomb one) can be considered
classical ones. For charmonium the part $n_{\rm coul}$ of gluon concentration 
$n$ connected with the Coulomb electric colour field is decreasing faster than 
$n_{\rm lin}$, 
the part of $n$ related to the linear magnetic colour field, and at large 
distances $n_{\rm lin}$ becomes dominant. Under the circumstances, as has been
said above, we can estimate the quark velocities in the charmonium 
state under discussion from the condition 
$$v=\frac{1}{\sqrt{1+\left(\frac{\lambda_B}{\lambda_q}\right)^2}}\>
\eqno(9.10)$$
with the $c$-quark Compton wavelength 
$\lambda_q=1/(2\mu_0)\approx0.168247\ {\rm fm}$ so, taking the de Broglie 
wavelength $\lambda_B=0.1r_1$ with $r_1$ from Table 3, 
we obtain $v\approx0.999718$, i. e., the quarks in
charmonium should be considered the ultrarelativistic point-like particles.
This additionally confirms the conclusion of Refs. \cite{{Gon03},{GC03}} that
the relativistic effects are extremely important for the confinement mechanism.
As a result, the confinement scenario described in Subsection 4.2 may 
really occur. At last, we can see that strength of magnetic colour field 
responsible for linear confinement reaches huge values of order 
$10^{17}$--$10^{18}\ {\rm T}$. For comparison one should notice that the most 
strong magnetic fields known at present have been discovered in magnetic 
neutron stars, pulsars (see, e. g., Ref. \cite{Bo}) where the corresponding 
strengths can be of order $10^{9}$--$10^{10}\ {\rm T}$. So the characterictic 
feature of confinement is really the very strong magnetic colour field between 
quarks which we have emphasized in Subsection 5.1. In a certain sense the 
essence of confinement can be said to be just in enormous gluon concentrations 
and magnetic colour field strentghs in space around quarks.

\section{Conclusion}
  Throughout the paper we moved step-by-step forward in analysing the 
Dirac-Yang-Mills system of equations (5.4)--(5.5) derived from the QCD 
lagrangian. The aim we pursued was to obtain a scenario for linear confinement 
of quarks, at any rate, in mesons and quarkonia. It seems to us we succeeded 
in finding the suitable quantitative description for the given phenomenon. As 
we could see, crucial step here consisted in studying exact solutions of the 
SU(3)-Yang-Mills equations modelling confinement and the corresponding modulo 
square integrable solutions of the Dirac equation. Techniques of finding the 
mentioned solutions were based to a great degree on those borrowed from black hole 
theory. In this respect our approach is conventional for physics whose whole history 
shows that very often the methods developed for solving some problems proved to be 
extremely useful for analysing a number of other ones in a perfectly different 
region of physics. 

 At the end of our considerations let us summarize the main features of the 
confinement mechanism developed in the given paper. 
\begin{enumerate}
\item  The whole approach is based on the exact solutions of the Yang-Mills 
equations and on the corresponding modulo square integrable solutions of Dirac 
equation. The solutions in question are essentially unique ones: for the 
confining solutions of the SU(3)-Yang-Mills equations that was shown in 
Section 8 whereas uniqueness for the conforming solutions of Dirac equation 
(5.4) follows from the ansatz $7.1^\prime$. Namely, the mentioned ansatz leads 
to (7.3) which is the confluent hypergeometric equation and (as is known from 
literature on special functions, see, e. g., Ref. \cite{Abr64}) the 
latter just possesses the only suitable solutions for us in the form of 
Laguerre polynomials which in essence determines the spectrum of relativistic 
bound states [see relations (7.7)--(7.8)]. Consequently, we found the only 
compatible solutions of the system (5.4)--(5.5) which can have pretensions of 
describing the confinement mechanism. Another matter that lagrangian QCD 
probably allows one to develop some other approaches 
[not based on compatible solutions of (5.4)--(5.5) ] to confinement but our 
one seems to be 
the most natural since it is practically identical to the standard approach 
of quantum mechanics and QED to description of bound states in external 
electromagnetic fields. In other words, this approach should have been one of 
the very first approaches as soon as the QCD lagrangian was written. Historically, 
however, this way was rejected due to incomprehensible reasons. 

\item  Two main physical reasons for linear confinement in the mechanism under 
discussion are the following ones. The first one is that gluon exchange between 
quarks is realized with the propagator different from the photon one and 
existence of such a propagator is direct sequence of the unique confining 
solutions of the Yang-Mills equations. The second reason is that, owing to 
the structure of mentioned propagator, gluon condensate (a classical gluon 
field) between quarks mainly consists of soft gluons ( see Subsection 4.1) but,  
because of that any gluon also emits gluons, the corresponding gluon concentrations 
rapidly become huge and form the linear confining magnetic colour field of 
enormous strengths which leads to confinement of quarks. Under the circumstances 
physically nonlinearity of the Yang-Mills equations effectively vanishes so the 
latter possess only the unique confining solutions of the abelian-like form (with the 
values in Cartan subalgebra) that describe the gluon condensate under 
consideration. Moreover, since the overwhelming majority of gluons are soft they 
cannot leave hadron (meson) until some gluon obtains additional energy (due to 
an external reason) to rush out.  The latter seems to be observable just in the 
so-called 3-jets events (for more details see, e. g., Ref. \cite{Per}). So we 
deal with confinement of gluons as well.

\item  The approach under discussion equips us with the explicit wave 
functions that is practically unreachable in other approaches, for example, 
within framework of lattice gauge theories or potenial approach. Namely, 
for each meson there exists its own set of real constants 
$a_j, A_j, b_j, B_j$ parametrizing the mentioned confining gluon
field (the gluon condensate) and the corresponding wave 
functions while the latter ones also depend on $\mu_0$, the reduced
mass of the current masses of quarks forming 
meson. It is clear that constants $a_j, A_j, b_j, B_j,\mu_0$
should be extracted from experimental data. This circumstance gives 
possibilities for direct physical modelling of internal structure for any meson 
and for checking such relativistic models numerically. 

\item Finally, there is also an interesting possibility of indirect experimental 
verification of the confinement mechanism under discussion. Really solutions 
(2.6) point out the confinement phase could be in electrodynamics as well. 
Though there exist no elementary charged particles generating a constant magnetic 
field linear in $r$, the distance from particle, after all, if it could generate 
this elecromagnetic field configuration in laboratory then one might study 
motion of trial charged particles in that field. The confining properties 
of the mentioned field should be displayed at classical level too but the exact 
behaviour of particles in this field requires certain analysis of the 
corresponding classical equations of motion.

\end{enumerate}
 
  At this point we would like to mark the end of the given paper on pinnig our 
hopes on studying confinement with subsequent development of the results 
obtained here, in the first place, by further applications to concrete mesons. 

%\newpage
\section{Appendix $A$: Hodge Star Operator $*$ on Minkowski \\ Spacetime 
in Spherical Coordinates}
Let $M$ is a smooth manifold of dimension $n$ so we denote an 
algebra of smooth functions on $M$ as $F(M)$. In a standard way the spaces 
of smooth differential $p$-forms $\Lambda^p(M)$ ($0\le p\le n$) are defined 
over $M$ as modules over $F(M)$ 
(see, e. g. Refs. \cite{Bes87}). If a (pseudo)riemannian metric 
$G=ds^2=g_{\mu\nu}dx^\mu\otimes dx^\nu$ is given on $M$ in local coordinates 
$x=(x^\mu)$ then $G$ can naturally be continued on spaces $\Lambda^p(M)$ 
by relation 
$$G(\alpha,\beta)={\rm det}\{G(\alpha_i,\beta_j)\}    \eqno(A.1)$$
for $\alpha=\alpha_1\wedge\alpha_2...\wedge\alpha_p$, 
$\beta=\beta_1\wedge\beta_2...\wedge\beta_p$, where for 1-forms
$\alpha_i=\alpha_\mu^{(i)}dx^\mu$, $\beta_j=\beta_\nu^{(j)}dx^\nu$ we have 
$G(\alpha_i,\beta_j)=g^{\mu\nu}\alpha_\mu^{(i)}\beta_\nu^{(j)}$ with the 
Cartan's wedge (external) product $\wedge$. Under the circumstances the Hodge 
star operator $*$: $\Lambda^p(M)\to\Lambda^{n-p}(M)$ is defined for any 
$\alpha\in\Lambda^p(M)$ by
$$\alpha\wedge(*\alpha)=G(\alpha,\alpha)\omega_g\> \eqno(A.2)$$
with the volume $n$-form 
$\omega_g=\sqrt{|\det(g_{\mu\nu})|}dx^1\wedge...dx^n$. 
For example, for 2-forms $F=F_{\mu\nu}dx^\mu\wedge dx^{\nu}$ we have
$$
F\wedge\ast F=(g^{\mu\alpha}g^{\nu\beta}-g^{\mu\beta}g^{\nu\alpha})
F_{\mu\nu}F_{\alpha\beta}
\sqrt{\delta}\,dx^1\wedge dx^2\cdots\wedge dx^n,\,\mu<\nu,\,\alpha<\beta 
\>\eqno(A.3)
$$
with $\delta=|\det(g_{\mu\nu})|$.
If $s$ is the number 
of (-1) in a canonical presentation of quadratic form $G$ then two most 
important properties of $*$ for us are 
$$ *^2=(-1)^{p(n-p)+s}\>,\eqno(A.4)$$
$$ *(f_1\alpha_1+f_2\alpha_2)=f_1(*\alpha_1)+f_2(*\alpha_2)\> \eqno(A.5)$$
for any $f_1, f_2 \in F(M)$, $\alpha_1, \alpha_2 \in\Lambda^p(M)$, i. e., 
$*$ is a $F(M)$-linear operator. Due to ($A.5$) for description 
of $*$-action in local coordinates it is enough to specify $*$-action on 
the basis elements of $\Lambda^p(M)$, i. e. on the forms 
$dx^{i_1}\wedge dx^{i_2}\wedge...\wedge dx^{i_p}$ with $i_1<i_2<...<i_p$ 
whose number is equal to $C_n^p=\frac{n!}{(n-p)!p!}$.

The most important case of $M$ in the given paper is the Minkowski spacetime 
with local coordinates $t, r, \vartheta, \varphi$, where 
$r, \vartheta, \varphi$ stand for spherical coordinates on spatial part of $M$. 
The metric is given by (1.1) and we shall obtain the $*$-action on the basis 
differential forms according to ($A.2$)
$$\ast dt=r^2\sin{\vartheta}dr\wedge d\vartheta\wedge d\varphi,\>
\ast dr=r^2\sin{\vartheta}dt\wedge d\vartheta\wedge d\varphi,\>$$
$$\ast d\vartheta=-r\sin{\vartheta}dt\wedge dr\wedge d\varphi,\>
\ast d\varphi=rdt\wedge dr\wedge d\vartheta,\>$$
$$\ast(dt\wedge dr)=-r^2\sin\vartheta d\vartheta\wedge d\varphi\>,
\ast(dt\wedge d\vartheta)=\sin\vartheta dr\wedge d\varphi\>,$$
$$\ast(dt\wedge d\varphi)=-\frac{1}{\sin\vartheta}dr\wedge d\vartheta\>,
\ast(dr\wedge d\vartheta)=\sin\vartheta dt\wedge d\varphi\>,$$
$$\ast(dr\wedge d\varphi)=-\frac{1}{\sin\vartheta}dt\wedge d\vartheta\>,
\ast(d\vartheta\wedge d\varphi)=\frac{1}{r^2\sin\vartheta}dt\wedge dr\>,$$
$$\ast(dt\wedge dr\wedge d\vartheta)=\frac{1}{r}d\varphi\>,
\ast(dt\wedge dr\wedge d\varphi)=-\frac{1}{r\sin{\vartheta}}d\vartheta,\>$$
$$\ast(dt\wedge d\vartheta\wedge d\varphi)=\frac{1}{r^2\sin{\vartheta}}dr,\>
\ast(dr\wedge d\vartheta\wedge d\varphi)=
\frac{1}{r^2\sin{\vartheta}}dt,\>
\eqno(A.6)$$
so that on 2-forms $\ast^2=-1$, as should be in accordance with 
($A.4$).

At last it should be noted that all the above is easily over linearity 
continued on the matrix-valued differential forms (see, e. g., Ref.\cite{Car}),  
i. e., on the arbitrary linear combinations of forms 
$a_{i_1i_2...i_p}dx^{i_1}\wedge dx^{i_2}\wedge...\wedge dx^{i_p}$, 
where coefficients $a_{i_1i_2...i_p}$ belong to some space of matrices $V$, 
for example, a SU($N$)-Lie algebra. In the latter case, if $T_a$ are 
matrices of 
generators of the SU($N$)-Lie algebra in $N$-dimensional representation, we 
continue the above scalar product $G$ on the SU($N$)-Lie algebra valued 1-forms 
$A=A^a_\mu T_adx^\mu$ and $B=B^b_\nu T_bdx^\nu$ by the relation
$$G(A,B)=g^{\mu\nu}A^a_\mu B^b_\nu{\rm Tr}(T_aT_b)\,,\eqno(A.7)$$
where Tr signifies the trace of a matrix, 
and on linearity with the help of ($A.1$), ($A.7$) can be continued 
over any SU($N$)-Lie algebra valued forms.

\section{Appendix $B$: SU($N$)-Lie Algebras and their Cartan \\Subalgebras}
\subsection*{SU($N$)-Yang-Mills Fields as Connections in Vector Bundles}
The most convenient mathematical treatment of classical SU($N$)-Yang-Mills 
fields is, to our mind, the one decribing those fields as connections in 
vector (or, which is equivalent, in principal) bundles over some manifold 
$M$. When $M$ is the Minkowski space the situation is simplified so long as 
all the bundles over $M$ with topology ${\Bbb R}^4$ are trivial. Referring for 
more details, e. g. to Refs. \cite{YM}, we shall restrict ourselves to a few 
remarks. 

If denoting $\xi$ the standard (trivial) $N$-dimensional vector bundle over 
Minkowski spacetime we can introduce an SU($N$)-connection (a classical 
SU($N$)-Yang-Mills field) in $\xi$ as 
a SU($N$)-Lie algebra valued form (the connection matrix) 
$A=A_\mu dx^\mu$, $A_\mu=A^a_\mu T_a$ while 
the matrices $T_a$ form
a basis of the Lie algebra of group SU($N$) in $N$-dimensional space (we consider
$T_a$ hermitian, as is accepted in physics), $a=1,...,N^2-1$. Then 
the curvature matrix (field strentgh)
for $\xi$-bundle is $F=dA+gA\wedge A=F^a_{\mu\nu}T_adx^\mu\wedge dx^\nu$
with the exterior differential $d$ (for example, 
$d=\partial_t dt+\partial_r dr+
\partial_\vartheta d\vartheta+\partial_\varphi d\varphi$ in coordinates
$t,r,\vartheta,\varphi$) while a gauge coupling constant $g$ is introduced 
from physical considerations.
Any smooth function (the gauge transformation) $S: M\to\,$SU($N$) gives the 
so-called trivialization of $\xi$-bundle 
and if $A_S$ and $F_S$ are the connection and curvature matrices, 
respectively, for the given trivialization then they are related to the 
previous ones by relations (the gauge transformations)
$$A_S=S^{-1}AS+\frac{i}{g}S^{-1}dS,\>
F_S=S^{-1}FS,\> \eqno(B.1)$$
where the factor $i$ is taken in order to an SU($N$)-Lie algebra could be chosen 
from the hermitian matrices.

Mathematically $A$ and $F$ are linked by the Bianchi identity holding true 
for any connection
$$dF=F\wedge A - A\wedge F \>, \eqno(B.2)$$
and physical considerations impose the Yang-Mills equations
$$d\ast F= g(\ast F\wedge A - A\wedge\ast F) +J\>,\eqno(B.3)$$
where $\ast$ means the Hodge star
operator conforming to a Minkowski metric, for instance, in the form of (1.1), 
while $J$ is a source, i. e., some differential SU($N$)-algebra valued 3-form. 
An arbitrary SU($N$)-connection does not obey the equations ($B.3$) except for
the so-called self-dual fields for those $F=\ast F$ and the equations 
($B.2$) and ($B.3$) become the same (at $J=0, g=1$). But, as was remarked 
in Appendix $A$, in Minkowski spacetime $\ast^2=-1$ on 2-forms that 
entails $F=0$ for self-dual fields. 

Also it should be noted that the solutions of ($B.3$) are usually 
believed to obey an additional condition.
In the present paper we take the Lorentz condition that can be written
in the form ${\rm div}(A)=0$, where the divergence of the Lie algebra valued
1-form $A=A^a_\mu T_adx^\mu$ is defined by the relation (see, e. g.
Refs. \cite{Bes87})
$${\rm div}(A)=\frac{1}{\sqrt{\delta}}\partial_\mu(\sqrt{\delta}g^{\mu\nu}
A_\nu)\>.\eqno(B.4)$$

It should be emphasized that the Lorentz condition is necessary for 
quantizing the gauge fields consistently within the framework of perturbation 
theory (see, e. g. Ref. \cite{Ryd85}), so we should impose the given condition.

At last, the case of group U(1) is also very important. We then have the case 
of classical electromagnetic field (the corresponding Lie algebra consists 
from real numbers), the Bianchi identity is converted into the the first pair 
of Maxwell equations $dF= 0$ while 
the Yang-Mills equations ($B.3$) become the second pair of Maxwell equations
$$d\ast F= J\>\eqno(B.5)$$
with $F=dA$, $A=A_\mu dx^\mu$ and $J=j_\mu*(dx^\mu)$ with a 4-dimensional
electromagnetic density current $j=j_\mu dx^\mu$, where $j_\mu=j_\mu(x)$ are 
some functions on ${\Bbb R}^4$.
 
Let us describe the explicit realizations of SU($N$)-Lie algebras 
by hermitian matrices that are
needed for our aims and also indicate the so-called 
Cartan subalgebras in them which is important in the main part of the paper. 
By definition, a Cartan subalgebra is a maximal abelian subalgebra in 
the corresponding Lie algebra, i. e., the commutator of any two matrices of 
the Cartan subalgebra is equal to zero. Dimension of Cartan subalgebra 
(as a vector space) for SU($N$)-Lie algebra is equal to $N-1$. 

\subsection*{$N=2$}
In this case we can take $T_a=\sigma_{a}$ at $a=1,2,3$, where
$\sigma_{a}$ are the ordinary Pauli matrices
$$\sigma_1=\pmatrix{0&1\cr 1&0\cr}\,,\sigma_2=\pmatrix{0&-i\cr i&0\cr}\,,
\sigma_3=\pmatrix{1&0\cr 0&-1\cr}\,\,. \eqno(B.6)$$
We shall also notice that 
$${\rm Tr}(\sigma_k\sigma_j)=\frac{1}{2}\delta_{kj}\,. \eqno(B.7)$$
The Cartan subalgebra is generated by $\sigma_3$.
\subsection*{$N=3$}
In the given situation we can take $T_a=\lambda_{a}$, 
where $\lambda_{a}$ are the Gell-Mann matrices
$$\lambda_1=\pmatrix{0&1&0\cr 1&0&0\cr 0&0&0\cr}\,,
  \lambda_2=\pmatrix{0&-i&0\cr i&0&0\cr 0&0&0\cr}\,,
  \lambda_3=\pmatrix{1&0&0\cr 0&-1&0\cr 0&0&0\cr}\,,  $$
$$ \lambda_4=\pmatrix{0&0&1\cr 0&0&0\cr 1&0&0 \cr}\,,
   \lambda_5=\pmatrix{0&0&-i\cr 0&0&0\cr i&0&0\cr}\,,
   \lambda_6=\pmatrix{0&0&0\cr 0&0&1\cr 0&1&0\cr}\,, $$
$$\lambda_7=\pmatrix{0&0&0\cr 0&0&-i\cr 0&i&0\cr}\,,
  \lambda_8={1\over\sqrt3}\pmatrix{1&0&0\cr 0&1&0\cr 
                   0&0&-2\cr}\,, \eqno(B.8)   $$
and 
$${\rm Tr}(\lambda_k\lambda_j)=2\delta_{kj}\,. \eqno(B.9)$$
The Cartan subalgebra is generated by $\lambda_3$ and $\lambda_8$.
\subsection*{$N=4$}
Following Ref. \cite{RF} we take ($I_2$ stands for the unit matrix 
$2\times2$)
$$T_1=\pmatrix{\sigma_1&0\cr 0&\sigma_1\cr}\,,
T_2=\pmatrix{\sigma_2&0\cr 0&\sigma_2\cr}\,,
T_3=\pmatrix{\sigma_3&0\cr 0&\sigma_3\cr}\,,$$
$$T_4=\pmatrix{0&I_2\cr I_2& 0\cr}\,,
T_5=\pmatrix{0&-iI_2\cr iI_2& 0\cr}\,,
T_6=\pmatrix{I_2 & 0\cr 0 & -I_2\cr}\, \eqno(B.10) $$
and the products of matrices $T_7=T_1T_4$, $T_8=T_1T_5$, $T_9=T_1T_6$, 
$T_{10}=T_2T_4$, $T_{11}=T_2T_5$, $T_{12}=T_2T_6$, $T_{13}=T_3T_4$, 
$T_{14}=T_3T_5$, $T_{15}=T_3T_6$ for the rest of 
the Lie algebra basis so the Cartan subalgebra is generated by 
$T_3, T_6, T_{15}$ with
$$T_{15}=T_3T_6=\pmatrix{\sigma_3 & 0\cr 0 & -\sigma_3\cr}\,.
\eqno(B.11)$$
Then
$${\rm Tr}(T_kT_j)=4\delta_{kj}\,. \eqno(B.12)$$

\section{Appendix $C$: Eigenspinors of the Euclidean Dirac Operator 
on ${\Bbb S}^2$ at $\lambda=\pm1$}
When using the wave functions obtained in Section 7 in any applications 
(see, e.g., Ref. \cite{Gon04}) one needs the explicit form for eigenspinors of 
the euclidean Dirac operator ${\cal D}_0$ (see Subsection 6.1). Though it is 
in general given by the relation (6.6), for applications, as a rule, it is 
sufficient to restrict oneself to small eigenvalues of ${\cal D}_0$. 
Let us write out, therefore, the eigenspinors of the euclidean Dirac operator 
${\cal D}_0$ corresponding to the eigenvalues $\lambda=\pm1$ in explicit form. 

If $\lambda=\pm(l+1)=\pm1$ then $l=0$ and from (6.6) it follows that 
$k=l+1/2=1/2$, $|m'|<1/2$ and we need the functions $P^{1/2}_{\pm1/2\pm1/2}$. 
But according to Ref. \cite{Vil91} we have $P^{k}_{kk}=
\cos^{2k}{(\vartheta/2)}$,  
$P^{k}_{k-k}=i^{2k}\sin^{2k}{(\vartheta/2)}$ and, besides, 
$P^{k}_{kk}=P^{k}_{-k-k}$, $P^{k}_{-kk}=P^{k}_{k-k}$ that entails the 
eigenspinors for $\lambda=-1$ in the form 
$$\Phi=\frac{C}{2}\pmatrix{\cos{\frac{\vartheta}{2}}+
i\sin{\frac{\vartheta}{2}}\cr
\cos{\frac{\vartheta}{2}}-i\sin{\frac{\vartheta}{2}}\cr}e^{i\varphi/2}, 
\Phi=\frac{C}{2}\pmatrix{\cos{\frac{\vartheta}{2}}+
i\sin{\frac{\vartheta}{2}}\cr
-\cos{\frac{\vartheta}{2}}+i\sin{\frac{\vartheta}{2}}\cr}
e^{-i\varphi/2},\eqno(C.1)$$
while for $\lambda=1$ the conforming spinors are
$$\Phi=\frac{C}{2}\pmatrix{\cos{\frac{\vartheta}{2}}-
i\sin{\frac{\vartheta}{2}}\cr
\cos{\frac{\vartheta}{2}}+i\sin{\frac{\vartheta}{2}}\cr}e^{i\varphi/2}, 
\Phi=\frac{C}{2}\pmatrix{-\cos{\frac{\vartheta}{2}}+
i\sin{\frac{\vartheta}{2}}\cr
\cos{\frac{\vartheta}{2}}+i\sin{\frac{\vartheta}{2}}\cr}e^{-i\varphi/2}
\eqno(C.2) $$
with the coefficient $C=\sqrt{1/(2\pi)}$.

\section{Appendix $D$: Condition $\sigma_2\Phi\approx\sin{\vartheta}\,\Phi$}

Let us compute spinor $\Phi_0=(\sigma_2-\sin{\vartheta})\,\Phi$ 
where, for example, $\Phi$ is the first spinor in ($C.1$). We obtain 
$$\Phi_0=\sqrt{\frac{1}{8\pi}}\left[-\sin{\frac{\vartheta}{2}}\left(2+
\cos{\vartheta}\right)\pmatrix{1\cr 1\cr}+
i\cos{\frac{\vartheta}{2}}\left(2-\cos{\vartheta}\right)
\pmatrix{-1\cr 1\cr}\right]e^{i\varphi/2}\>, 
\eqno(D.1)$$
so that both components of $\Phi_0$ are modulo 
$$\sqrt{\frac{1+3\sin^2{\vartheta}}{8\pi}}\le\sqrt{\frac{1}{2\pi}}
\approx0.4<1\>,$$
i. e. condition $\sigma_2\Phi\approx\sin{\vartheta}\,\Phi$ is good enough 
fulfilled. This holds true for all the eigenspinors of ${\cal D}_0$, or, 
more generally, of ${\cal D}_n$ which 
we shall not dwell upon.

\section{Appendix $E$: More General Ansatz} 

As was remarked in Subsection 8.1, there can be a more general ansatz than 
(8.1) for finding the confining solutions, namely it is 
$$A=r^{\mu_a}\alpha^a\lambda_a dt+A_rdr+
A_\vartheta d\vartheta+
r^{\nu_a}\beta^a\lambda_a d\varphi\>,\eqno(E.1)$$ 
with $A_\vartheta=r^{\rho_a}\gamma^a\lambda_a$ and arbitrary real constants 
$\alpha^a, \beta^a, \gamma^a$. 
Under this situation the Lorentz condition ($B.4$) for the given ansatz entails 
$\cot{\vartheta}r^{\rho_a}\gamma^a\lambda_a+\partial_r(r^2A_r)=0$ wherefrom 
it follows 
$$A_r=\frac{C}{r^2}-\frac{\cot{\vartheta}r^{\rho_a-1}}{\rho_a+1}
\gamma^a\lambda_a\>\eqno(E.2)$$
with a constant matrix $C$ belonging to SU(3)-Lie algebra. 
Then we can see that it should put $C=\gamma^a=0$ or else $A_r$ will not be 
spherically symmetric and the confining one where only the powers of $r$ equal 
to $\pm1$ are admissible. As a result, we come to the conclusion that one 
should put 
$A_r=A_\vartheta=0$ in ($E.1$).

After this we have 
$$dA= -\mu_ar^{\mu_a-1}\alpha^a\lambda_a dt\wedge dr+
\nu_ar^{\nu_a-1}\beta^a\lambda_a dr\wedge d\varphi,$$ 
$$F=dA+gA\wedge A=dA+g\alpha^a\beta^br^{\mu_a+\nu_b}
[\lambda_a,\lambda_b]dt\wedge d\varphi\>\eqno(E.3)$$
and, with the help of ($A.6$), we obtain 
$$\ast F=\mu_ar^{\mu_a+1}\alpha^a\lambda_a\sin{\vartheta}
d\vartheta\wedge d\varphi-\frac{ 
\nu_ar^{\nu_a-1}\beta^a\lambda_a}{\sin{\vartheta}}
dt\wedge d\vartheta-$$
$$gr^{\mu_a+\nu_b}\alpha^a\beta^b\frac{
[\lambda_a,\lambda_b]}{\sin{\vartheta}}dr\wedge d\vartheta  
\>, \eqno(E.4)$$
$$d\ast F=\mu_a(\mu_a+1)r^{\mu_a}\alpha^a\lambda_a\sin{\vartheta}
dr\wedge d\vartheta\wedge d\varphi+\frac{ 
\nu_a(\nu_a-1)r^{\nu_a}\beta^a\lambda_a}{\sin{\vartheta}}
dt\wedge dr\wedge d\vartheta \>,\eqno(E.5)$$
so the Yang-Mills equations ($B.3$) (at $J=0$) turn into the system 
$$\mu_a(\mu_a+1)r^{\mu_a}\alpha^a\lambda_a\sin^2{\vartheta}=
-g^2r^{\mu_a+\nu_b+\nu_c}\alpha^a\beta^b\beta^c
[[\lambda_a,\lambda_b],\lambda_c]\>,
               \eqno(E.6)$$
$$\nu_a(\nu_a-1)r^{\nu_a-2}\beta^a\lambda_a=
-g^2r^{\mu_a+\nu_b+\mu_c}\alpha^a\beta^b\alpha^c
[[\lambda_a,\lambda_b],\lambda_c]\>,
               \eqno(E.7)$$
$$g(\mu_a-\mu_b)r^{\mu_a+\mu_b}\alpha^a\alpha^b
[\lambda_a,\lambda_b]=0\>,\>
g(\nu_a-\nu_b)r^{\nu_a+\nu_b}\beta^a\beta^b
[\lambda_a,\lambda_b]=0\>, a<b,\>. 
               \eqno(E.8)$$
Then, obviously, in the left-hand sides of ($E.6$--$E.7$) we should put all 
$\mu_a=-1$ and all $\nu_a=1$, respectively, for obtaining the confining 
solutions while the right-hand sides of ($E.6$--$E.7$) are identically equal 
to zero only if $\lambda_{a,b,c}=\lambda_{3,8}$. Under the circumstances the 
equations ($E.8$) are satisfied automatically and we come to the same 
conclusions about uniqueness of the confining solutions as in Subsection 8.1. 

%\newpage

\end{document}